\documentclass[11pt,draftcls,peerreview,onecolumn,twoside]{IEEEtran}
\usepackage[T1]{fontenc}
\usepackage{amsmath}
\usepackage{graphicx}
\usepackage{pb-diagram}
\usepackage{amssymb}

\newtheorem{definitn}{Definition}
\newtheorem{prop}{Proposition}
\newtheorem{thm}{Theorem}
\newtheorem{cor}{Corollary}
\newtheorem{example}{Example}
\newtheorem{remark}{Remark}
\usepackage{verbatim}
\newtheorem{lemma}{Lemma}
\newtheorem{remrk}{Remark}

\newcommand{\ZZ}{\mathbb{Z}}
\newcommand{\QQ}{\mathbb{Q}}
\newcommand{\CC}{\mathbb{C}}

\newcommand{\Oc}{\mathcal{O}}
\newcommand{\A}{\mathcal{A}}
\newcommand{\Ic}{\mathcal{I}}
\newcommand{\Cc}{\mathcal{C}}
\newcommand{\Nc}{\mathcal{N}}

\newcommand{\Xm}{\mathbf{X}}
\newcommand{\Rm}{\mathbf{R}}
\newcommand{\zm}{\mathbf{0}}
\newcommand{\vv}{\mathbf{v}}
\newcommand{\Gal}{\mbox{Gal}}
\newcommand{\Tr}{\mbox{Tr}}

\begin{document}

\title{Perfect Space Time Block Codes}

\author{Frédérique Oggier, Ghaya Rekaya, \IEEEmembership{Student Member~IEEE},
Jean-Claude Belfiore, \IEEEmembership{Member~IEEE} and Emanuele Viterbo,
\IEEEmembership{Member~IEEE}
\thanks{Frédérique Oggier is with the California Institute of
  Technology, 91125 Pasadena, California, USA. When this work was
  done, she was with \'Ecole Polytechnique Fédérale de Lausanne,
1015 Lausanne, SWITZERLAND}
\thanks{Ghaya Rekaya and Jean-Claude Belfiore are with \'Ecole Nationale
Supérieure des Télécommunications, 46 rue Barrault, 75013 Paris, FRANCE }
\thanks{Emanuele Viterbo is with Politecnico di Torino, C.so Duca degli
Abruzzi 24, 10129 Torino, ITALY }
\thanks{Emails: {\tt frederique@systems.caltech.edu,
\{rekaya,belfiore\}@enst.fr,viterbo@polito.it}}}

\maketitle

\markboth{Submitted to IEEE Trans. IT}{Oggier \emph{et al.} Perfect STBCs
...}

\begin{abstract}
In this paper, we introduce the notion of perfect space-time block codes
(STBC).
These codes have full rate, full diversity, non-vanishing constant
minimum determinant for increasing spectral efficiency, uniform average
transmitted energy per antenna and good shaping. We present algebraic
constructions of perfect STBCs for $2$, $3$, $4$ and $6$ antennas.
\end{abstract}
\IEEEpeerreviewmaketitle

%
%

\section{Introduction}

\PARstart{I}n order to achieve very high spectral efficiency over
wireless channels, it is known that we need multiple antennas at both
transmitter and receiver ends. We consider the coherent case where
the receiver has perfect knowledge of all the channel coefficients.
It has been shown \cite{Tarokh-1} that the main code design criterion
in this scenario is the {\em rank criterion}: the rank of the difference of
two distinct codewords has to be maximal. If this property is satisfied,
the codebook is said to be fully diverse. Once the difference has full
rank, the product of its singular values is nonzero, and is defining
the {\em coding gain}. Maximizing the coding gain is the second design
criterion. Extensive work has been
done on designing Space-Time codes that are fully diverse.

We focus here on Linear Dispersion Space-Time Block Codes (LD-STBC),
introduced in \cite{Hassibi-2}. The idea of LD codes is to spread
the information symbols over space and time. The linearity property
of the LD-STBC enables the use of maximum likelihood (ML) sphere
decoding \cite{Viterbo99,Hassibi01}, which exploits the full
performance of the code compared to other suboptimal decoders
\cite{Damen-1}. Consequently, research work has been done to
construct LD-STBCs with more structure. One new property added has
been {\em full rate}, i.e., the number of transmitted signals
corresponds to the number of information symbols to be sent, in
order to maximize the throughput. In \cite{Damen-3}, it is shown how
to construct full rate and fully diverse codes for the $2$ transmit
antennas case. This approach is generalized for any number $M$ of
transmit antennas in \cite{El-Gamal-2,Galliou-2}. A promising
alternative approach based on {\em division algebras} is proposed in
\cite{Rajan-03}, where the authors construct non-full-rate and
full-rate STBCs. A division algebra (as it will be detailed below)
is an algebraic object that naturally yields a linear set of
invertible matrices. It can thus be used to construct LD codes,
since for any codeword the rank criterion is satisfied.

In \cite{Belfiore-3,Belfiore-2}, we have presented the Golden code,
a $2\times2$ STBC obtained using a division algebra, which is full
rate, full diversity, and has a nonzero lower bound on its coding
gain, which does not depend on the constellation size. A code
isomorphic to the Golden code was independently found by an
analytical optimization in \cite{Yao-1} and \cite{Varanasi-2}. In
\cite[Theorem 1]{Yao-1}, it is also shown that, for $2$ antennas, a
sufficient condition for achieving the diversity-multiplexing gain
frontier defined by Zheng and Tse \cite{Zheng} is exactly the lower
bound on the coding gain. In \cite{Elia}, it has been shown in
general that the nonzero lower bound on the coding gain is actually
a sufficient condition to reach the frontier for any number of
antennas.

The goal of this work is to refine the code design criteria for
LD-STBCs, asking for the three following properties:
\begin{itemize}
\item A nonzero lower bound on the coding gain, which is independant
of the spectral efficiency ({\em non-vanishing determinant}).
\item What we call a {\em shaping constraint}, to guarantee that
the codes are energy efficient.
\item Uniform average transmitted energy per
antenna is also required.
\end{itemize}
We propose the so-called {\em perfect codes} that fulfill the above
properties, and give explicit constructions in dimension $2,3,4$ and
$6$ for $2\times 2$, $3\times 3$, $4\times 4$ and $6\times 6$ MIMO
systems.

The paper is organized as follows.
In Section \ref{sec:probl_stat}, we detail the code design criteria and
define precisely the notion of {\em perfect codes}.
Since our code constructions are based on cyclic algebras, we begin
Section \ref{sec:cycl_alg} by recalling how one can use cyclic
division algebras to build fully-diverse and full-rate STBCs.
We then explain further algebraic techniques useful to obtain the
properties of the perfect codess. The following parts of the
paper are dedicated to the code constructions.
In Section \ref{sec:inf_two}, we exhibit an infinite family of
$2\times2$ perfect STBCs generalizing the Golden Code construction
\cite{Belfiore-2}.
Then, we construct a $3\times3$, a $4\times4$ and a $6\times6$ perfect
STBC in Sections resp.~\ref{sec_3}, \ref{sec_4} and \ref{sec_6}.

%
%

\section{Problem statement}\label{sec:probl_stat}

We consider a coherent system over a flat fading $M\times N$
MIMO channel, where the receiver knows the channel coefficients
(perfect CSI). The received matrix is
\begin{equation}
\mathbf{Y}_{N\times T}=\mathbf{H}_{N\times M}\cdot\mathbf{X}_{M\times T}
                      +\mathbf{W}_{N\times T},
\label{eq:schem-tran}
\end{equation}
where $\mathbf{X}$ is the transmitted codeword of duration $T$
taken from a STBC $\Cc$, $\mathbf{H}$ is the channel matrix
with i.i.d. Gaussian entries and $\mathbf{W}$
is the i.i.d. Gaussian noise matrix. Subscripts indicate the dimensions
of the matrices.

In this paper, we consider square ($M=T$) linear dispersion
STBCs \cite{Hassibi-2} with full-rate
i.e., square codes with $M$ degrees of freedom, using either QAM or
HEX \cite{Forney-3} information symbols.
Since the codewords are square, we can reformulate the rank criterion
saying that the codebook is fully diverse if
\[
|\det(\Xm_i-\Xm_j)|^2 \neq 0,~\Xm_i \neq \Xm_j \in \Cc.
\]
By linearity, this simplifies to
$|\det(\Xm)|^2\neq 0$, for all nonzero codeword $ \Xm \in \Cc$.

Once a codebook is fully-diverse, the next step attempts
to maximize the coding advantage, which is defined for LD-STBC by
the \textit{minimum determinant} of the code.
We first consider infinite codes defined by assuming that the information
symbols are allowed to take values in an infinite constellation.
The \textit{minimum determinant} of the infinite code $\Cc_{\infty}$ is
\[
\delta_{\min}(\Cc_{\infty})
=\min_{\zm \neq \Xm \in\Cc_{\infty}}\left|\det(\mathbf{X})\right|^{2}.
\]
We denote by $\Cc$ the finite code obtained
by restricting the information symbols to $q\textrm{-QAM}$ constellations
or $q\textrm{-HEX}$.
The \textit{minimum determinant} of $\Cc$ is then
\[
\delta_{\min}(\mathcal{C})
=\min_{\zm \neq \Xm \in\Cc}|\det(\Xm)|^{2}.
\]
%
In \cite{Rajan-03} as well as in all the previous constructions
\cite{El-Gamal-2,Galliou-2}, the emphasis is on having
a non-zero minimum determinant. But since the minimum determinant is
dependent on the spectral efficiency, it vanishes when the constellation
size increases.

{\bf Non-vanishing determinant.}
We say that a code has a {\em non-vanishing determinant} if, without
power normalization, there is a lower bound on the minimum determinant
that does not depend on the constellation size. In other words,
we impose that the minimum determinant of the STBC is a constant
$\Delta_{\min }$ for a sufficiently high spectral efficiency.
For low spectral efficiencies, it is lower-bounded by $\Delta_{\min }$.
Non-vanishing determinants may be of interest,
whenever we want to apply some outer block coded modulation scheme,
which usually entails a signal set expansion, if the spectral efficiency
has to be preserved.

A fixed minimum determinant is one of the two key properties of the
\emph{perfect codes} introduced in this work, the other one is related
to the constellation shaping.

{\bf Shaping.}
In order to optimize the energy efficiency of the codes,
we introduce a shaping constraint on the signal constellation.
It is enough to introduce this shaping constraint on each layer as
the codes considered in this paper all use the layered structure
of \cite{Hammons-1}. The $q$-QAM or $q$-HEX to be sent are normalized
according to the power at the transmitter. However, since we use
LD-STBCs, what is transmitted on each layer is not just information
symbols but a linear combination of them, which may change the
energy of the signal. Each layer can be written as $\Rm \vv$,
where $\vv$ is the vector containing the
QAM or HEX information symbols, while $\Rm$ is a matrix that
encodes the symbols into each layer. In order to get energy efficient
codes, we ask the matrix $\Rm$ to be unitary.
We will refer to this type of constellation shaping as
{\em cubic shaping}, since a unitary matrix applied on a vector
containing discrete values can be interpreted as generating points in
a lattice. For example, if we use QAM symbols, we get the $\ZZ^n$
(cubic) lattice.

The last property of perfect codes is related to the energy per antenna.

{\bf Uniform average energy transmitted per antenna.} The $i$th antenna of
the system will transmit the $i$th row of the codeword. We ask that on
average, the norm of each row are similar, in order to have a balanced
repartition of the energy at the transmitter.
It was noticed in \cite{Rajan-03} that uniform average
transmitted energy per antenna in all $T$ time slots is required.

We are now able to give the definition of a \emph{perfect} STBC code.
\begin{definitn}
A square $M\times M$ STBC is called a \emph{perfect} code if and only if:
\end{definitn}
\begin{itemize}
\item
It is a full rate linear dispersion code using $M^{2}$ information
symbols either QAM or HEX.
\item
The minimum determinant of the infinite code is non zero (so that
in particular the rank criterion is satisfied).
\item
The energy required to send the linear combination of the information
symbols on each layer is similar to the energy used for sending the
symbols themselves (we do not increase the energy of the system in
encoding the information symbols).
\item
It induces uniform average transmitted energy per antenna
in all $T$ time slots, i.e., all
the coded symbols in the code matrix have the same average energy.
\end{itemize}

Let us illustrate the definition by showing that the Golden code, the
$2\times2$ STBC presented in \cite{Belfiore-2} is a perfect STBC.

\begin{example}
A codeword $\mathbf{X}$ belonging to the Golden Code has the form
\[
\mathbf{X} =
 \frac{1}{\sqrt{5}}\left[\begin{array}{cc}
\alpha(a+b\theta) & \alpha(c+d\theta)\\
i\bar{\alpha}(c+d\bar{\theta}) & \bar{\alpha}(a+b\bar{\theta})
\end{array}\right]
\]
where $a,b,c,d$ are QAM symbols,
$\theta=\frac{1+\sqrt{5}}{2}$,
$\bar\theta=\frac{1-\sqrt{5}}{2}$, $\alpha=1+i-i\theta$ and
$\bar{\alpha}=1+i -i\bar{\theta}$.

The code is full rate since it contains 4 information symbols,
$a,b,c,d$.
Let us now compute the minimum determinant of the infinite code.
Since $\alpha\bar{\alpha}=2+i$, we have
\begin{eqnarray*}
\det(\Xm) & = & \frac{2+i}{5}[(a+b\theta)(a+b\bar{\theta})
                             -i(c+d\theta)(c+d\bar{\theta})] \\
          & = & \frac{1}{2-i}[(a^2+ab-b^2-i(c^2+cd-d^2)].
\end{eqnarray*}
By definition of $a,b,c,d$, we have that the minimum of
$|a^2+ab-b^2-i(c^2+cd-d^2)|^2$ is 1, thus
\[
\delta_{\min}(\Cc_{\infty})=\min_{\zm \neq \Xm \in\Cc}|\det(\Xm)|^2
= \frac{1}{5}.
\]
Thus the minimum determinant of the infinite code is bounded away from
zero, as required.

Let us now consider the diagonal layer of the code. It can be written
\[
\frac{1}{\sqrt{5}}
\left(
\begin{array}{cc}
\alpha & \alpha \theta \\
\bar{\alpha} & \bar{\alpha \theta}\\
\end{array}
\right)
\left(
\begin{array}{c}
a \\
b\\
\end{array}
\right).
\]
Since the matrix can be checked to be unitary, the cubic shaping is
satisfied.

Note in the second row of the codeword $\Xm$ the factor $i$,
which guarantees uniform average transmitted energy since $|i|^2=1$.
\end{example}

This code has of course been designed to satisfy all the required
properties. Its main structure comes from a division algebra, and
the shaping is obtained by interpreting the signals on each layer as
points in a lattice. In the following, we explain the algebraic tools
we use, and show how to obtain codes with similar properties for
a larger number of antennas.

%
%

\section{Cyclic algebras: a tool for space-time coding}\label{sec:cycl_alg}

We start by recalling the most relevant concepts about cyclic algebras
and how to use them to build full rate and fully diverse space-time
block codes (see also \cite{Rajan-03} for more details ). We then
explain how to add more structure on the algebra to
get the other properties required to get perfect codes, namely, the
shaping constraint and the non-vanishing determinant.
We warn the reader that some algebraic background is
required. If the reader is not familiar with the notions of norm,
trace, Galois group, or discriminant, we recommand to read first the
appendix \ref{app:basics} where these notions are recalled.


\subsection{Full rate and fully diverse STBCs}

In the following, we consider number field extensions $K/F$, where $F$
denotes the base field.
The set of non-zero elements of $F$ (resp. $K$) is denoted by $F^*$
(resp. $K^*$).

Let $K/F$ be a cyclic extension of degree $n$, with Galois group
$\Gal(K/F)=\langle\sigma\rangle$, where $\sigma$ is the generator of the cyclic
group. Let $\A=(K/F,\sigma,\gamma)$ be its corresponding
{\em cyclic algebra} of degree $n$, that is
\[
\A = 1\cdot K \oplus e\cdot K \oplus \ldots \oplus e^{n-1} \cdot K
\]
with $e\in\A$ such that $l e = e \sigma(l)$ for all $l\in K$
and $e^{n}=\gamma\in F^*$.

Cyclic algebras provide families of matrices by
associating to an element $x \in \A$ the matrix of multiplication by $x$.
\begin{example}\rm
For $n=2$, we have $\A= 1\cdot K \oplus e \cdot K$ with $e^2 = \gamma$ and
$l e = e \sigma(l)$ for $l \in K$.
An element $x \in \A$ can be written $x=x_0+ex_1$.
Let us compute the multiplication by $x$ of any element $y \in \A$.
\begin{eqnarray*}
xy &= & (x_0+ex_1)(y_0+ey_1) \\
   &= & x_0y_0 + e\sigma(x_0)y_1 +ex_1y_0+ \gamma \sigma(x_1)y_1 \\
   &= & [x_0y_0+ \gamma \sigma(x_1)y_1]+e[\sigma(x_0)y_1 +x_1y_0],
\end{eqnarray*}
since $e^2=\gamma$ and using the noncommutativity rule
$l e = e \sigma(l)$.

\noindent In the basis $\{1,e\}$, this yields
\[
xy=
\left(
\begin{array}{cc}
x_0 & \gamma \sigma(x_1) \\
x_1 & \sigma(x_0) \\
\end{array}
\right)
\left(
\begin{array}{c}
y_0 \\
y_1  \\
\end{array}
\right).
\]
There is thus a correspondance
\[
x=x_0+ex_1 \in \A \leftrightarrow
\left(
\begin{array}{cc}
x_0 & \gamma \sigma(x_1) \\
x_1 & \sigma(x_0) \\
\end{array}
\right).
\]
In particular,
\[
e \in \A \leftrightarrow
\left(
\begin{array}{cc}
0 & \gamma \\
1 &  0 \\
\end{array}
\right).
\]
\end{example}
In the general case of degree $n$, we have for all $x_k \in K$
\[
x_k \leftrightarrow
\left( \begin{array}{cccc}
        x_k & 0            &        & 0  \\
         0  & \sigma(x_k)  &        & 0  \\
    \vdots  &              & \ddots & \vdots  \\
         0  &    0         &        & \sigma^{n-1}(x_k) \\
       \end{array}
\right) ~~\mbox{ and }~~ e \leftrightarrow \left(
\begin{array}{cccccc}
         0      &  0  &  0      &     &     & \gamma \\
         1      &  0  &  0      &     &     &   0    \\
         0      &     & \ddots  &     &     &        \\
         0      &     &         &     & 1   &  0    \\
        \end{array} \right).
\]
Formally, one can associate a matrix to any element $x\in\A$ using
the map $\lambda_{x}$, the multiplication by $x$ of an element $y\in\A$:
\[
\begin{array}{rcl}
\lambda_{x}:\A & \rightarrow & \A\\
y & \mapsto & \lambda_{x}(y)=x\cdot y.\end{array}
\]
The matrix of the multiplication by $\lambda_{x}$, with
$x=x_0+ e x_1 +\ldots+ e^{n-1} x_{n-1}$,
is more generally given by
\begin{equation}
\left(\begin{array}{ccccc}
x_0 & \gamma\sigma(x_{n-1}) & \gamma\sigma^2(x_{n-2}) & \ldots & \gamma\sigma^{n-1}(x_1)    \\
x_1 & \sigma(x_0)           & \gamma\sigma^2(x_{n-1}) & \ldots & \gamma\sigma^{n-1}(x_2) \\
\vdots  &                   & \vdots                  &        & \vdots                  \\
x_{n-2} & \sigma(x_{n-3})   & \sigma^2(x_{n-4})       & \ldots & \gamma\sigma^{n-2}(x_{n-1})\\
x_{n-1} & \sigma(x_{n-2}) & \sigma^2(x_{n-3})  & \ldots & \sigma^{n-1}(x_0)
\end{array}\right).
\label{eq:matrix}
\end{equation}
Thus, via $\lambda_x$, we have a {\em matrix representation} of an element
$x \in \A$.

Let us show how encoding can be done. All the coefficients of such
matrices are in $K$, $K$ being a vector space of dimension $n$ over
$F$. Thus each $x_i$ is a linear combination of $n$ elements in $F$.
The information symbols are thus chosen to be in $F$.
If we consider QAM constellations with in-phase and quadrature
$\pm 1, \pm3 ,\ldots$, the constellation can be seen as a subset of
$\ZZ[i]:=\{a+bi,~a,b \in \ZZ\}$ (Gaussian integers). Since
$\ZZ[i]\subset \QQ(i)$, we take $F=\QQ(i)$ in order to transmit
$q$-QAM. Similarly, in order to use HEX symbols, we see them as
a subset of $\ZZ[j]:=\{a+bj,~a,b \in \ZZ\}$ (Eisenstein integers)
where $j$ is a primitive 3rd root of unity ($j^3=1$, $j=e^{2i\pi/3}$).
We then take $F=\QQ(j)$ with $\ZZ[j] \subset \QQ(j)$. Following the
terminology of \cite{Rajan-03}, we may say that the STBC
$\Cc_{\infty}$ is \emph{over} $F$.

The following space-time block code is then obtained
\begin{equation}\label{eq:Cc}
\Cc_{\infty}=\left\{ \left(\begin{array}{cccc}
x_0 & x_1 & \ldots & x_{n-1}\\
\gamma\sigma(x_{n-1}) & \sigma(x_0) & \ldots & \sigma(x_{n-2})\\
\vdots &  &  & \vdots\\
\gamma\sigma^{n-1}(x_{1}) & \gamma\sigma^{n-1}(x_{2}) & \ldots & \sigma^{n-1}(x_{0})\end{array}\right)~|~x_{i}\in K,~i=0,\ldots,n-1\right\} .
\end{equation}
Since each codeword $\Xm$ contains $n$ coefficients $x_i$, each of
them being a linear combination of $n$ information symbols, cyclic
algebras naturally yields {\em full rate} LD-STBCs.

\begin{definitn}\label{def:norm_red}
The determinant of the matrix (\ref{eq:matrix}) (which is also the
determinant of a codeword (\ref{eq:Cc})) is called the
\emph{reduced norm} of $x$, $x \in \mathcal{A}$.
\end{definitn}

The key point of this algebraic scheme is that we have a criterion to decide
whether the STBC $\Cc_{\infty}$ satisfies the rank criterion.
Namely, when the cyclic algebra is a division algebra, all its elements
are invertible; hence the codeword matrices have non zero determinants.
\begin{prop}
\label{pro:norm}\cite{Rajan-03} The algebra $\A=(K/F,\sigma,\gamma)$ of
degree $n$ is a division algebra if the smallest positive integer $t$ such
that $\gamma^{t}$ is the norm of some element in $K^{*}$ is $n$.
\end{prop}
So at this point, by choosing an element $\gamma$ such that its powers
are not a norm, the codebook $\Cc_{\infty}$ defined in (\ref{eq:Cc})
is a fully diverse LD-STBC with full rate.


\subsection{The shaping constraint using complex algebraic lattices}
\label{sub:Complex-algebraic-lattices}

The shaping constraint requires that each layer of the codeword is
of the form $\Rm\vv$, where $\Rm$ is a unitary matrix and $\vv$ is a
vector containing the information symbols. Let $K=F(\theta)$ and
$\{1,\theta,\ldots,\theta^{n-1}\}$ be a $F$-basis of $K$. Each layer
of a codeword $\Xm$ as in (\ref{eq:Cc}) is of the form
\begin{equation}\label{eq:layer}
\left(
\begin{array}{cccc}
1 & \theta & \ldots & \theta^{n-1}\\
1 & \sigma(\theta) & \ldots & \sigma(\theta^{n-1})\\
  &                &        &  \\
1 & \sigma^{n-1}(\theta) & \ldots & \sigma^{n-1}(\theta^{n-1})\\
\end{array}
\right)
\left(
\begin{array}{c}
u_{l,0} \\
u_{l,1}\\
\vdots\\
u_{l,n-1}
\end{array}
\right)
=
\left(
\begin{array}{c}
x_l \\
\sigma(x_l)\\
\vdots\\
\sigma^{n-1}(x_l)
\end{array}
\right)
\end{equation}
for $x_l=\sum_{k=0}^{n-1}u_{l,k}\theta^k$.
Since $u_{l,k}$ takes discrete values, we can see the above matrix
multiplication as generating points in a lattice. The matrix $\Rm$ is
thus the {\em generator matrix} of the lattice, and the lattice
obtained is given by $\Rm\Rm^H$, its {\em Gram matrix}.
We would like $\Rm$ to be unitary, which translates into saying
that the lattice we would like to obtain for each layer is  a
$\ZZ[i]^n$--lattice, resp. a $\ZZ[j]^n$--lattice, since QAM and HEX
symbols as finite subsets of $\ZZ[i]$, resp. $\ZZ[j]$.
Note that the matrix $\Rm$ may be viewed as a precoding matrix
applied to the information symbols.

Finally, note that the $2n^2$--dimensional real lattice generated by the
vectorized codewords where real and imaginary components are separated,
is either $\ZZ^{2 n^2}$ (for QAM constellation) or
$A_2^{n^2}$ (for HEX constellation), where $A_2$ is the
hexagonal lattice \cite{splag}, with generator matrix
\[
\left(\begin{array}{cc}
1 & 0 \\
1/2 & \sqrt{3}/2
\end{array}
\right).
\]

Interpreting the unitary matrix $\Rm$ as the generator matrix of a
lattice allows us to use the well studied theory of {\em algebraic lattices}
\cite{Bayer-1,Bayer-2,FnT}. The key idea is that the matrix $\Rm$
given in (\ref{eq:layer}) needs to contain the embeddings of a basis,
but this basis does not need to be a basis of the field $K$. It can be
a basis of a subset of $K$, and in fact it will be a basis of
an {\em ideal} of $K$.

Let $K$ be a Galois extension of $F=\QQ(i)$ (resp. $F=\QQ(j)$) of
degree $n$, and denote by $\Oc_{K}$ its ring of integers.
Let $\QQ(\theta)$ be a totally real Galois number field of degree $n$
with discriminant coprime to the one of $F$, that is
$(d_{F},d_{\QQ(\theta)})=1$.
In the following, we focus on the case where $K$ is the compositum of
$F$ and $\QQ(\theta)$ (that is, the smallest field that contains both).
We write the compositum as $K=F\QQ(\theta)$ (see Fig.~\ref{fig:comp}).
This assumption has the convenient consequence that \cite[p.~48]{Swinnerton}
\begin{equation}\label{eq:coprime_dk}
d_K =   d_{\QQ(\theta)}^2 d_{F}^n,
\end{equation}
where $d_{F}=-4$ for $F=\QQ(i)$ and $d_{F}=-3$ for $F=\QQ(j)$.

Denote by $\{\sigma_{k}\}_{k=1}^{n}$ the Galois group Gal($K/F$).
\begin{definitn}\label{def:compl_latt}
We denote by $\Lambda^c(\Ic)$ the {\em complex algebraic lattice}
corresponding to an ideal $\Ic\subseteq\Oc_K$
obtained by the complex embedding $\sigma$ of $K$ into
$\CC^{n}$ defined as
\begin{align*}
\hbox{\boldmath$\sigma$}: &~ K\rightarrow\CC^{n}\\
 & x\mapsto \hbox{\boldmath$\sigma$}(x)=(\sigma_{1}(x),\ldots,\sigma_{n}(x))
\end{align*}
\end{definitn}
The basis of $\Lambda^c(\Ic)$ is obtained by embedding the
basis $\{\nu_{k}\}_{k=1}^{n}$of $\Ic$. Consequently its generator
matrix is similar to the matrix $\Rm$ in (\ref{eq:layer}), where the
basis $\{1,\theta,\ldots,\theta^{n-1}\}$ is replaced by the ideal basis
$\{\nu_{k}\}_{k=1}^{n}$. Its Gram matrix $G$ is thus given by
\[
G = (\Tr_{K/F}(\nu_{k}\overline{\nu_{l}}))_{k,l=1}^{n},
\]
where $\overline{x}$ denotes the complex conjugation of $x$.
When $F=\QQ(j)$, since
Gal($\QQ(j)/\QQ$)=$\langle \tau \rangle$, with $\tau(j)=j^{2}=\bar{j}$,
we have that $\tau$ coincides with the complex conjugation.

We explain now how to choose an ideal $\Ic \subseteq \Oc_K$ in order to get
the rotated versions of the $\ZZ^{2n}$ or $A_2^{n}$ lattices.
First consider the real lattice
$\Lambda(\Ic)$ obtained from $\Lambda^c(\Ic)$ by vectorizing
the real and imaginary parts of the complex lattice vectors.
We want   $\Lambda(\Ic)$ to be a rotated version of $\ZZ^{2n}$ or $A_2^n$.
The basic idea is that the norm of the ideal $\Ic$ is closely
related to the volume of $\Lambda(\Ic)$.
We will thus look for an ideal with the ``right'' norm.

\begin{itemize}
\item
Consider the {\em ramification} in $K/\QQ$, that is the way prime
numbers in $\ZZ$ may factorize when considered in $K$ (for example,
5 is prime in $\ZZ$ but is not prime anymore in $\QQ(i)$, since
$5=(2+i)(2-i)$). We say that a prime $p_k$ {\em ramifies} if
$(p_{k})\Oc_{K}=\prod_{\ell}\mathcal{I}_{k\ell}^{e_{k}}$ where
$e_{k}>1$ \cite[p.~86]{Samuel} for some $k$ (or in words, the primes
which when factorized in $K$ have factors with a power greater or
equal to 2). The prime factorization of the discriminant
$d_{K/\QQ}=\prod p_{k}^{r_{k}}$ contains the primes $p_k$ which
ramify \cite[p.~88]{Samuel}.
\item
Considering real algebraic lattices $\Lambda(\Oc_{K})$
\cite{Bayer-1}, we know that
vol$(\Lambda(\Oc_{K}))=2^{-n}\sqrt{|d_{K/\QQ}|}$.
We look for a sublattice $\Lambda(\Ic)$ of $\Lambda(\Oc_{K})$, which
could be a scaled version of $\ZZ^{2n}$ (resp. $A_{2}^{n}$),
i.e., $\Lambda(\Ic)=(\sqrt{c}\ZZ)^{2n}$
(resp. $\left(cA_{2}\right)^{n}$) for some integer $c$.
\item
Since $\Lambda(\Ic)$ is a sublattice of $\Lambda(\Oc_{K})$,
vol$(\Lambda(\Oc_{K}))=2^{-n}\sqrt{|d_{K}|}$
must divide
\[
\textrm{vol}(\Lambda(\Ic))=
\left\{ \begin{array}{l}
        \textrm{vol}\left((\sqrt{c}\ZZ)^{2n}\right)=c^n\\
        \textrm{vol}\left((cA_2)^n\right)=c^{n}\left(\frac{\sqrt{3}}{2}\right)^{n}
        \end{array}
\right.
\]
i.e., $d_{K/\QQ}=\prod p_{k}^{r_{k}}$ divides $2^{2n}c^{2n}$ (resp. $3^{n}c^{2n}$).
\item
This gives a necessary condition for the choice of $\Ic$. In terms
of norm of the ideal $\Ic$ \cite[p.~69]{Samuel}, we need
\begin{equation}
N(\Ic)=|\Oc_{K}/\Ic|
=\frac{\mbox{vol}(\Lambda(\Ic))}{\mbox{vol}(\Lambda(\Oc_{K}))}
=\left\{ \begin{array}{c}
         \frac{(2c)^{n}}{\sqrt{\prod p_{k}^{r_{k}}}}\\
         \frac{\left(\sqrt{3}c\right)^{n}}{\sqrt{\prod p_{k}^{r_{k}}}}
         \end{array}\right.\label{eq:norm_ideal}
\end{equation}
Recall from (\ref{eq:coprime_dk}) that $d_{K}=2^{2n}\cdot d^2_{\QQ(\theta)}$,
when $K$ is the compositum of $\QQ(i)$ and $\QQ(\theta)$ with coprime
discriminants and that $d_{K}=3^{n}\cdot d^2_{\QQ(\theta)}$,
when $K$ is the compositum of $\QQ(j)$ and $\QQ(\theta)$ with coprime
discriminants.
\item
In order to satisfy (\ref{eq:norm_ideal}), we must find an ideal
$\Ic$ with norm $\prod_{p_{k}\neq 2} p_{k}^{n-r_{k}/2}$
(resp. $\prod_{p_{k}\neq3}p_{k}^{n-r_{k}/2}$).
\end{itemize}

This procedure helps us in guessing what is the ``right'' ideal
$\Ic$ to take in order to build a $\ZZ^{2n}$ or $A_2^n$ lattice.
To prove that we indeed found the ``right'' lattice it is sufficient to
show that
\begin{equation}
\label{eq:trace_delta}
\Tr_{K/F}(\nu_i \bar{\nu_j})= \delta_{i,j},~i,j =1,\ldots,n
\end{equation}
where $\{\nu_i\}_{i=1}^n$ denotes the basis of the ideal $\Ic$, and
$\delta_{i,j}$ is the Kronecker delta.

Note that the lattice does not exist on all field extensions $K/F$.
Once we have a cyclic field extension where the lattice exists, we define
a fully diverse full rate codebook which furthermore satisfies the
shaping constraint as
\begin{equation}\label{eq:CI}
\Cc_{\Ic}=
\left\{
\left(\begin{array}{cccc}
x_0 & x_1 & \ldots & x_{n-1}\\
\gamma\sigma(x_{n-1}) & \sigma(x_0) & \ldots & \sigma(x_{n-2})\\
\vdots &  &  & \vdots\\
\gamma\sigma^{n-1}(x_1) & \gamma\sigma^{n-1}(x_2) & \ldots &
\sigma^{n-1}(x_0)
\end{array}\right)
~|~x_i\in\Ic\subseteq\Oc_K,~i=0,\ldots,n-1
\right\}.
\end{equation}


\subsection{Discreteness of the determinants}\label{subsec:discrete_det}

The goal of this section is to show how to get codes built over a
cyclic algebra $\A=(K/F, \sigma,\gamma)$ so that their determinants
are discrete. One condition will appear to be $\gamma \in \Oc_F$,
the ring of integer of $F$. This contrasts with the approach of
Sethuraman {\em et al.} \cite[Proposition 12]{Rajan-03}, where the
element $\gamma$ was chosen to be transcendental. Hence, the cyclic
division algebra $(K(\gamma)/F(\gamma),\sigma,\gamma)$ is used,
which ensures that the minimum determinant is non-zero.
Unfortunately, this approach yields a vanishing minimum determinant,
when the constellation size increases.

In \cite{Belfiore-1}, it has been shown for $2\times2$ STBCs, by an
explicit computation of the determinant, that the reduced norm of the
algebra (see Def.~\ref{def:norm_red}) is linked to the algebraic norm of
elements in $K$. Since the norm of an element in $K$ belongs to $F$,
restricting the codeword matrix elements to be in $\Oc_{K}$
and taking $\gamma\in \Oc_F$ then gives discrete values of the
determinants for the codewords of $2\times2$ STBCs.
The same result has also been used in \cite{Belfiore-2}, for the
Golden code. However, an explicit determinant computation is no more
possible in higher dimensions. We thus invoke a general result that
guarantees the reduced norm to be in $F$.

\begin{thm}\label{thm:Scharlau}\cite[p. 296  and p. 316]{Scharlau}
Let $\A=(K/F,\sigma,\gamma)$
be a cyclic algebra, then its reduced norm belongs to $F$.
\end{thm}
\begin{cor}\label{cor:Scharlau}
Let $\A=(K/F,\sigma,\gamma)$ be a cyclic algebra with $\gamma \in \Oc_F$.
Denote its  basis by $\{1,e,\ldots,e^{n-1}\}$.
Let $x\in\A$ be of the form
\[
x=x_0+ e x_1 +\ldots+ e^{n-1} x_{n-1}
\]
where $x_{k}\in\Oc_{K}$, $k=0,\ldots,n-1$.
Then, the reduced norm of $x$ belongs to $\Oc_{F}$.
\end{cor}
\begin{proof}
Recall from Definition \ref{def:norm_red} that the reduced norm of $x$ is the
determinant of its matrix representation. Since $x_k \in \Oc_K$ implies
$\sigma(x_k) \in \Oc_K$ for all $k$ and $\gamma \in \Oc_F$ by
hypothesis, all coefficients of the matrix representation belong to
$\Oc_{K}$, hence so does its determinant.
By Theorem \ref{thm:Scharlau}, the reduced norm of $x$ belongs to $F$, so
it belongs to $\Oc_{K}\cap F=\Oc_{F}$.
\end{proof}

\begin{cor} \label{cor:detmineq1}
The minimum determinant of the infinite code with $\Ic=\Oc_K$
defined in (\ref{eq:CI}) is
\[
\delta_{\min}(\Cc_{\Oc_K})=1.
\]
\end{cor}
\begin{proof}
Since we only consider $\Oc_{F}=\ZZ[i]$ (resp.
$\Oc_{F}=\ZZ[j]$), the determinants of the codewords
form a discrete subset of $\CC$:
\[
\det({\bf X}) \in \ZZ[i] ~~(\mbox{resp.}~ \in\ZZ[j]).
\]
Then $\delta_{\min}(\Cc_{\Oc_K}) =
\min_{{\bf X}\neq{\bf 0} } |\det({\bf X})|^2 = 1$
as the minimum is achieved by taking the codeword with
$x_0=1$ and $x_k=0$ for $k=1\ldots n-1$, corresponding to a
single information symbol $u_{00}=1$ and all the
remaining $n^2-1$ equal to 0.
\end{proof}
Let us give as example the $3\times3$ case to show that things
become more complicated than the $2\times2$ case when the dimension
increases, so that the general Theorem \ref{thm:Scharlau} is required.
\begin{example}\label{ex:dim3}
Consider a cyclic algebra $\A=(K/F,\sigma,\gamma)$ of degree 3 with
$\gamma \in \Oc_F$.
Let $x=x_0+ e x_1 +e^2 x_2$, which can be represented as
\[
{\bf X}=\left[\begin{array}{ccc}
x_0 & x_1 & x_2\\
\gamma\sigma(x_2) & \sigma(x_0)& \sigma(x_1)\\
\gamma\sigma^2(x_1) & \gamma\sigma^2(x_2) & \sigma^2(x_0)
\end{array}\right].
\]
The norm of $x$ is given by the determinant of ${\bf X}$:
\begin{align*}
\det({\bf X}) = &\gamma^{2}x_{2}\sigma(x_{2})\sigma^{2}(x_{2})+\\
 & \gamma\{-x_{0}\sigma(x_{1})\sigma^{2}(x_{2})-
   \sigma(x_{0})\sigma^{2}(x_{1})x_{2}-
   \sigma^{2}(x_{0})x_{1}\sigma(x_{2})+
   x_{1}\sigma(x_{1})\sigma^{2}(x_{1})\}\\
 & +x_{0}\sigma(x_{0})\sigma^{2}(x_{0})\\
 = &N(x_{0})+\gamma N(x_{1})+\gamma^{2}N(x_{2})-
     \gamma\Tr[x_{0}\sigma(x_{1})\sigma^{2}(x_{2})]
\end{align*}
Obviously the norm of the algebra is not only related
to the norm of the number field, as in dimension 2
where $\det({\bf X})= N(x_{1})-\gamma N(x_{2})$ \cite{Belfiore-2}.
\end{example}

When considering a particular case, it is possible to conclude that
$\det({\bf X})$ still belongs to $F$, either as in Example \ref{ex:dim3}
by finding an expression in terms of norms and traces, or by noticing that
the determinant is invariant under the action of $\sigma$.
Since the expression in larger dimensions gets more complicated, for the
general case, we simply use Theorem \ref{thm:Scharlau}.

Note that at this point, we have all the ingredients to
build perfect codes. Assume there exists $\gamma \in \Oc_F$ such
that none of its powers is a norm. Then the code $\Cc_{\Ic}$ defined
in (\ref{eq:CI}) is fully diverse and full rate, it has the required
shaping constraint, and we have just shown that its determinant is
discrete. In order to conclude, it is now enough to take
$|\gamma|^2=1$, to guarantee uniform average transmitted energy
per antennas. Before summarizing our approach, we now give an
explicit bound on the minimum determinant.


\subsection{The minimum determinant}

We discuss now the value of the minimum determinant of the codes.
Depending on whether the ideal $\Ic$ introduced in subsection
\ref{sub:Our-approach} is principal (i.e., generated by one element),
we distinguish two cases.
We show that if $\Ic$ is principal, then the minimum determinant of the
infinite space-time code $\Cc_{\Ic}$ is easily computed. Otherwise, we
give a lower bound on $\delta_{\min}(\Cc_{\Ic})$.

Let us first assume $\Ic = (\alpha)\Oc_K$ is a principal ideal of $\Oc_K$.
For all $x \in \Ic$, we have $x= \alpha y$ for some $y \in \Oc_K$.
Notice that in this case, codewords are of the form
\begin{equation}
\mathbf{X}   =
\left[\begin{array}{cccc}
\alpha & 0 & \cdots & 0\\
0 & \sigma(\alpha) & \ddots & \vdots\\
\vdots & \ddots & \ddots & 0\\
0 & \cdots & 0 & \sigma^{n-1}(\alpha)
\end{array}\right]
\cdot
\left[\begin{array}{cccc}
y_{0} & y_{1} & \ldots & y_{n-1}\\
\gamma\sigma(y_{n-1}) & \sigma(y_{0}) & \ldots & \sigma(y_{n-2})\\
\vdots &  &  & \vdots\\
\gamma\sigma^{n-1}(y_{1}) & \gamma\sigma^{n-1}(y_{2}) & \ldots & \sigma^{n-1}(y_{0})
\end{array}\right]
\end{equation}
where $y_{i}\in\Oc_{K},\,\, i=0,\ldots,n-1$. Since $\gamma \in \Oc_F$,
the determinant
of the second matrix is in $\Oc_{F}$ and by Corollary \ref{cor:detmineq1}
its square modulus is at least 1.
We deduce, recalling that $F=\QQ(i)$ or $\QQ(j)$, that
\begin{equation}
\delta_{\min}(\Cc_{\Ic})
=\min_{\zm \neq \Xm\in\Cc_{\Ic}}\left|\det({\bf X})\right|^{2}
= |N_{K/F}(\alpha)|^{2} = N_{K/\QQ}(\alpha).
\label{eq:min-det-principal}
\end{equation}
The last equality is true since the complex conjugation is the
Galois group of $F/\QQ$. Thus
\[
|N_{K/F}(\alpha)|^2=\prod_{k=0}^{n-1}\sigma^k(\alpha)
   \prod_{k=0}^{n-1}\overline{\sigma^k(\alpha)}
\]
where $\sigma^k$, and  $\bar{\sigma}^k$, $k=0,\ldots,n-1$, give
the $2n$ elements of the Galois group of $K/\QQ$.

Since $K$ is the compositum of $F$ and a totally real
field  $\QQ(\theta)$ and we require the cubic shaping,
we can go a little further.
\begin{prop}\label{prop:dmin_principal}
Let $\Cc_{\Ic}$ be a perfect code built over the cyclic division algebra
$\A=(K/F,\sigma,\gamma)$ of degree $n$ where $\gamma \in \Oc_F$,
$K=F\QQ(\theta)$ and $\Ic$ is principal. Then
\[
\delta_{\min}(\Cc_{\Ic})=\frac{1}{d_{\QQ(\theta)}},
\]
where $d_{\QQ(\theta)}$ is the absolute discriminant of $\QQ(\theta)$.
\end{prop}
\begin{proof}
Let $\{\nu_i \}_{i=1}^n$ be a basis of the principal ideal
$\Ic=(\alpha) {\cal O}_K $ and $\Lambda(\Ic)$ denote the real lattice over $\ZZ$.
Recall \cite{Bayer-1} that
\begin{equation}\label{eq:det_norm_dk}
\det(\Lambda(\Ic)) = \mbox{vol}(\Lambda(\Ic))^2 =4^{-n} N(\Ic)^2 d_{K}
\end{equation}
where $d_{K}$ denotes the absolute discriminant of $K$.
Using (\ref{eq:coprime_dk}) and considering the real lattice, we have
for $F=\QQ(i)$
\[
\det(\ZZ^{2n}) = 1 = 4^{-n} N_{K/\QQ}(\alpha)^2 d_{\QQ(\theta)}^2 4^n,
\]
and for $F=\QQ(j)$
\[
\det(A_2^n) = (3/4)^n = 4^{-n} N_{K/\QQ}(\alpha)^2 d_{\QQ(\theta)}^2 3^n.
\]
Both cases reduce to
\[
N_{K/\QQ}(\alpha)= \frac{1}{d_{\QQ(\theta)}},
\]
and we conclude using (\ref{eq:min-det-principal}).
\end{proof}

We consider now the more general case, where we make no assumption on
whether $\Ic$ is principal. We have the following result.
\begin{prop}\label{prop:dmin_general}
Let $\Cc_{\Ic}$ be a perfect code built over the cyclic division algebra
$\A=(K/F,\sigma,\gamma)$ of degree $n$ where $\gamma \in \Oc_F$. Then
\[
\delta_{\min}(\Cc_{\Ic}) \in N(\Ic)\ZZ,
\]
where $N(\Ic)$ denotes the norm of $\Ic$.
\end{prop}
\begin{proof}
Recall first that
\[
\det({\bf X})=\sum_{s \in S_n}\mbox{sgn}(s)\prod_{k=1}^n ({\bf X})_{k,s(k)},
\]
where $S_n$ is the group of permutations of $n$ elements, and sgn denotes the
signature of the permutation.
Denote by $\Ic^{\sigma}$ the action of the Galois group on $\Ic$.
Since $({\bf X})_{k,s(k)} \in \Ic^{\sigma^{k-1}}$ for all $k$, we get
\cite[p.~118]{Frohlich}
\[
\det({\bf X}) \in \prod_{\sigma \in \Gal(K/F)}\Ic^{\sigma}
               = \Nc_{K/F}(\Ic)\Oc_K,
\]
where $\Nc_{K/F}(\Ic)$ stands for an ideal of $\Oc_F$ called the
{\em relative norm} of the ideal $\Ic$. The notation $\Nc(\Ic)$
emphasizes the fact that in the case of the relative norm of an ideal,
we deal with an ideal, and not with a scalar, as it is the case for
the absolute norm $N(\Ic)$ of an ideal.

Note that the above equation means that $\det(\Xm)$ belongs to an
ideal of $\Oc_K$. By Corollary \ref{cor:Scharlau}, we deduce that
\[
\det({\bf X}) \in \Oc_F \cap \Nc_{K/F}(\Ic)\Oc_K = \Nc_{K/F}(\Ic),
\]
which means that $\det(\Xm)$ is actually in an ideal of $\Oc_F$.
Thus $|\det({\bf X})|^2 \in \Nc_{F/\QQ}(\Nc_{K/F}(\Ic))$, since again
$F=\QQ(i)$ or $\QQ(j)$. We conclude
using the transitivity of the norm \cite[p.~99]{Frohlich}
\[
\min_{{\bf X}\in\Cc_{\Ic},{\bf X}\neq 0}\left|\det({\bf X})\right|^{2}
\in \Nc_{K/\QQ}(\Ic) = N(\Ic)\ZZ.
\]
\end{proof}
Bounds on $\delta_{\min}(\Cc_{\Ic})$ are easily derived from the above
proposition.
\begin{cor}\label{cor:bound_dmin}
Let $\Cc_{\Ic}$ be a perfect code built over the cyclic division algebra
$\A=(K/F,\sigma,\gamma)$ of degree $n$ where $\gamma \in \Oc_F$ and
$K=F\QQ(\theta)$. Then
\[
N(\Ic) = \frac{1}{d_{\QQ(\theta)}}
\leq
\delta_{\min}(\Cc_{\Ic})
\leq
\frac{1}{\mbox{vol}(\Lambda^c(\Ic))}\min_{x \in \Ic}N_{K/\QQ}(x)
\]
\end{cor}
\begin{proof}
The lower bound is immediate from Proposition \ref{prop:dmin_general} and
the equality comes from (\ref{eq:det_norm_dk}), similarly as in the
proof of Proposition \ref{prop:dmin_principal}.

An upper bound can be obtained as follows. We
take $x_0 \neq 0 \in \Ic$, $x_1 = \ldots = x_{n-1}= 0$, which yields
as determinant $N_{K/\QQ}(x_0)$. Thus
$\min \det(\Xm)= \min_{x \in \Ic}N_{K/\QQ}(x)$. Since the ideal $\Ic$
may give a scaled version of the lattice $\ZZ[i]^n$
(resp. $\ZZ[j]^n$), a normalizing factor given by the volume of the
lattice is necessary to make sure we consider a lattice with volume 1.
\end{proof}
The result obtained in (\ref{eq:min-det-principal}) for
the principal case alternatively follows:
\begin{cor}
If $\Ic=(\alpha)\Oc_K$ is principal, then
\[
\delta_{\min}(\Cc_{\Ic}) = N_{K/\QQ}(\alpha).
\]
\end{cor}
\begin{proof}
If $\Ic$ is principal, the lower and upper bounds in Corollary
\ref{cor:bound_dmin} coincide.
\end{proof}
%

\subsection{Summary of our approach\label{sub:Our-approach}}

Let us summarize the techniques explained above, and give the steps
we will follow in the next sections to construct perfect codes:
\begin{enumerate}
\item
We consider QAM or HEX symbols
with arbitrary spectral efficiency. Since these constellations can
be seen as finite subsets of the ring of integers $\Oc_F=\ZZ[i]$ (resp. $\Oc_F=\ZZ[j]$),
we take as base field $F=\QQ(i)$ (resp. $F=\QQ(j)$).
\item
We take a cyclic extension $K/F$ of degree $n=M$ with Galois group
$\Gal(K/F)=\langle\sigma\rangle$ and build the corresponding cyclic algebra:
\[
\A=(K/F,\sigma,\gamma).
\]
We choose $\gamma$ such that $|\gamma|=1$ in order to satisfy
the constraint on the uniform average transmitted energy per antenna.
\item
In order to obtain non-vanishing determinants, we choose
$\gamma$ in $\ZZ[i]$, resp. in $\ZZ[j]$
(see Sec.~\ref{subsec:discrete_det}). Adding the previous constraint
$|\gamma|=1$, we are limited to $\gamma \in \{1,i,-1,-i\} \subset \ZZ[i]$
or $\gamma \in \{1,j,j^2,-1,-j,-j^2\} \subset \ZZ[j]$, respectively.
\item
Among all elements of $\A$, we consider the discrete set of codewords
of the form $x=x_0+x_1 e+\ldots+x_{n-1} e^{n-1}$, where
$x_{i}\in\Ic$, an ideal of $\Oc_K$, the ring of integers of $K$.
This restriction on the coefficients guarantees a
discrete minimum determinant (see Section \ref{subsec:discrete_det}).
We thus get a STBC of the form
\begin{equation}
\Cc_{\Ic}=\left\{ \left(\begin{array}{cccc}
x_{0} & x_{1} & \ldots & x_{n-1}\\
\gamma\sigma(x_{n-1}) & \sigma(x_{0}) & \ldots & \sigma(x_{n-2})\\
\vdots &  &  & \vdots\\
\gamma\sigma^{n-1}(x_{1}) & \gamma\sigma^{n-1}(x_{2}) & \ldots & \sigma^{n-1}(x_{0})\end{array}\right)~|~x_{i}\in\Ic\subseteq\Oc_K,~
i=0,\ldots,n-1\right\}
\end{equation}
The $n^2$ information symbols $u_{\ell,k} \in \Oc_F$ are
encoded into codewords by
\[
x_\ell = \sum_{k=0}^{n-1} u_{\ell,k} \nu_k ~~~~~\ell=0,\ldots, n-1
\]
where $\{\nu_k\}_{k=0}^{n-1}$ is a basis of the ideal $\Ic$.
\item
We make sure to choose an ideal $\Ic \subseteq \Oc_K$ so that the
signal constellation
on each layer is a finite subset of the rotated versions of the lattices
$\ZZ^{2n}$ or $A_2^n$.
\item
We show that $\A=(K/F,\sigma,\gamma)$ is a division algebra
by selecting the right $\gamma$ among the possible choices which
reduces to show that $\gamma,\ldots,\gamma^{n-1}$ are not a norm in $K^{*}$.
\end{enumerate}
Since the desired lattice does not always exist, we need to choose
an appropriate field extension $K$ that gives both the lattice and a
division algebra. Note that, in building a cyclic algebra
$\A=(K/F,\sigma,\gamma)$ for STBCs, the choice of $\gamma$ is
critical since it determines whether $\A$ is a division algebra. It
is furthermore constrained by the requirement that $|\gamma|=1$, so
that the average transmitted energy by each antenna in all time
slots is equalized, and to be in $\Oc_F$ to ensure the discreteness
of the determinant.

\begin{remark}\rm
The construction of the codes involves a lot of computations in number fields.
Some of them are done by hand, some of them are computed with the
computational algebraic software Kant \cite{kant}.
\end{remark}

%
%

\section{An infinite family of codes for $2 \times 2$ MIMO}
\label{sec:inf_two}

In this section, we generalize the construction given in \cite{Belfiore-2}
to an infinite family of codes for $2\times 2$ MIMO.

Let $p$ be a prime. Let $K/\QQ(i)$ be a relative extension of degree
2 of $\QQ(i)$ of the form $K=\QQ(i,\sqrt{p})$. We can represent $K$
as a vector space over $\QQ(i)$:
\[
K=\{ a+b\sqrt{p}~|~a,b\in\QQ(i)\}.
\]
Its Galois group Gal$(K/\QQ(i))=\langle\sigma\rangle$ is generated by
$\sigma:\sqrt{p}\mapsto-\sqrt{p}$.
The corresponding cyclic algebra of degree 2 is
$\A=(K/\QQ(i),\sigma,\gamma)$.

We prove here that when $p\equiv5$ (mod 8), $\gamma=i$, and using
a suitable ideal $\Ic \subseteq \Oc_K$, we obtain perfect codes following
the scheme of Sec.~\ref{sub:Our-approach}.


\subsection{The lattice $\ZZ[i]^{2}$}

We first search for an ideal $\Ic \subseteq \Oc_K$ giving the rotated
$\ZZ[i]^{2}$ lattice.
We use the fact that $\mathbb{Z}[i]^{2}$ is the only \emph{unimodular}
$\mathbb{Z}[i]$--lattice in dimension 2 \cite{Schiemann}.
Hence it is enough to find an ideal $\mathcal{I}$ such that
$\Lambda^c(\mathcal{I})$ is unimodular.
By definition, a unimodular lattice coincides with its \emph{dual}
defined as follows.
Let $\Lambda^c(\mathcal{I})$ be a complex algebraic lattice with basis
$\{\mathbf{v}_{1},\mathbf{v}_{2}\}
=\{\mbox{\hbox{\boldmath$\sigma$}}(\nu_{1}),
   \mbox{\hbox{\boldmath$\sigma$}} (\nu_{2})\} $
following the notations of Section \ref{sub:Complex-algebraic-lattices}.

\begin{definitn}
The dual lattice of $\Lambda^c(\mathcal{I})$ is defined by
\[
\Lambda^c(\Ic)^{\#}=
\left\{ \mathbf{x}=a_{1}\mathbf{v}_{1}+a_{2}\mathbf{v}_{2},
a_{1},a_{2}\in\mathbb{Q}(i)|\,
\langle\mathbf{x},\mathbf{y}\rangle \in\mathbb{Z}[i],
\forall \mathbf{y}\in\Lambda^c(\mathcal{I})\right\}
\]
where the scalar product between the two vectors
can be related to the trace of the corresponding
algebraic numbers as
\[
\langle\mathbf{x},\mathbf{y}\rangle =
   \Tr_{K/\mathbb{Q}(i)}\left(x \overline{y}\right).
\]
\end{definitn}
The dual of a complex algebraic lattice can be computed explicitly.
Recall that the codifferent \cite[p. 44]{Swinnerton},\cite{Bayer-1}
is defined as
\[
D_{K/F}^{-1}=\{x \in K~|~\forall \alpha \in \Oc_K,
\mbox{Tr}_{K/F}(x\alpha) \in \Oc_F\}.
\]
\begin{lemma}
\label{lem:dual}
We have $\Lambda^c(\Ic)^{\#}=\Lambda^c\left(\Ic^{\#}\right)$ with
\[
\Ic^{\#}=\overline{\Ic^{-1}\mathcal{D}_{K/\QQ(i)}^{-1}}
\]
where $\mathcal{D}_{K/\QQ(i)}^{-1}$ denotes the codifferent (defined above).
\end{lemma}
\begin{proof}
Let $x\in\overline{\Ic^{-1}\mathcal{D}_{K/\QQ(i)}^{-1}}$. For all $y\in\Ic$,
we have to show that $\Tr_{K/\QQ(i)}(x\overline{y})\in\ZZ[i]$. Since
$x=\overline{uv}$, with $u\in\Ic^{-1}$ and $v\in\mathcal{D}_{K/\QQ(i)}^{-1}$,
we have $x\overline{y}=\overline{uy}\overline{v}$, with $uy\in\Oc_{K}$.
The result follows now from the definition of $\mathcal{D}_{K/\QQ(i)}^{-1}$.
\end{proof}
Let $K=\QQ(i,\sqrt{p})$, with $p\equiv1$ (mod 4). The factorization
of $p$ in $\Oc_{K}$ is \cite{Cohn}
\begin{equation}
(p)\Oc_{K}=\mathcal{I}^{2}\cdot\overline{\mathcal{I}}^{2}\label{eq:factor-p}\end{equation}
 where $\mathcal{I}$, $\overline{\mathcal{I}}$ are prime conjugate
ideals.

\begin{prop}
The $\mathbb{Z}[i]$--lattice $\frac{1}{\sqrt{p}}\Lambda^c(\Ic)$
is unimodular.
\end{prop}
\begin{proof}
Note first that $\mathcal{D}_{K/\QQ(i)}=\mathcal{D}_{\QQ(\sqrt{p})/\QQ}=(\sqrt{p})\Oc_{\QQ\left(\sqrt{p}\right)}=(\sqrt{p})$.
Using Lemma \ref{lem:dual} and (\ref{eq:factor-p}), we compute
the dual of $\Ic$,
\[
\Ic^{\#}=\overline{\Ic^{-1}}(\sqrt{p})^{-1}=\frac{1}{p}\Ic.
\]
Now the dual lattice is
\[
\left(\frac{1}{\sqrt{p}}\Lambda^c(\Ic)\right)^{\#}
=\sqrt{p}\left(\Lambda^c(\Ic)^{\#}\right)
=\frac{1}{\sqrt{p}}\Lambda^c(\Ic)
\]
which concludes the proof.
\end{proof}


\subsection{The norm condition}

The last step is to prove that the algebra $\A=(K/\QQ(i),\sigma,i)$
is a division algebra. In order to do that, we have to show (see Proposition
\ref{pro:norm}) that $\gamma=i$ is not a norm in $K/\QQ(i)$.

We first recall the characterization of a square in finite fields. Let $p$ be
a prime and denote by $GF(p)$ the finite field with $p$ elements.

\begin{prop}\label{pro:square}
Let $x\in GF(p)^{*}$. We have
\[
x\textrm{\, is a square\,$\Longleftrightarrow x^{\frac{p-1}{2}}=1$}.
\]
\end{prop}
\begin{proof}
See \cite{GF}.
\end{proof}
\begin{cor}
\label{cor:square}If $p\equiv1\textrm{\,(mod\,4)}$ , $-1$ is a
square in $GF(p)$.
\end{cor}
Let us come back to our case where $p$ is a prime such that $p\equiv5$
(mod 8) and $K=\QQ(i,\sqrt{p})$ is a relative extension of $\QQ(i)$.
Let $x=a+b\sqrt{p} \in K$, $a,b\in\QQ(i)$. Its relative norm is
\begin{equation}
N_{K/\QQ(i)}(x)=(a+b\sqrt{p})(a-b\sqrt{p})=a^{2}-pb^{2}.
\label{eq:norm_gen}
\end{equation}
Our goal is to show that the equation $N_{K/\QQ(i)}(x)=i$ has no
solution. As in \cite{Belfiore-2}, we prove that this equation has
no solution in the field of $p$-adic numbers $\QQ_{p}$, and thus, no
solution for $x\in K$.
Let $\ZZ_{p}=\{ x\in\QQ_{p}|\nu_{p}(x)\geq0\}$
be the valuation ring of $\QQ_{p}$, where $\nu_p(x)$ denotes the
{\em valuation} of $x$ in $p$ (that is, the power at which $p$ appears
in the factorization of $x$). First, we check that $i\in\ZZ_{p}$.
In fact, there are embeddings of $\QQ(i)$ into $\QQ_{p}$ if $X^{2}+1$,
the minimal polynomial of $i$, has roots in $\ZZ_{p}$. Using Hensel's
Lemma \cite[p.75]{Gouvea}, it is enough to check that -1 is a square
in $GF(p)$. By assumption, $p\equiv5$ (mod 8),
thus $p\equiv1\textrm{\,(mod\,4)}$,
then, by Corollary \ref{cor:square}, -1 is a square in $GF(p)$.

\begin{prop} \label{pro:not-norm}
The unit $i\in\mathbb{Z}[i]$ is not a relative
norm, i.e., there is no $x\in K$ such that $N_{K/\QQ(i)}(x)=i$ where
$K=\mathbb{Q}(\sqrt{p},i)$ with $p\equiv5$~(mod 8).
\end{prop}
\begin{proof}
This is equivalent, by (\ref{eq:norm_gen}), to prove that \begin{equation}
a^{2}-pb^{2}=i,~a,b\in\QQ(i)\label{eq:Q(i)_gen}\end{equation}
 has no solution.
 Using the embedding of $\QQ(i)$ into $\QQ_{p}$, this equation can
be seen in $\QQ_{p}$ as follows: \begin{equation}
a^{2}-pb^{2}=y+px,~a,b\in\QQ_{p},~x,y\in\ZZ_{p},\label{eq:Qp}\end{equation}
 where $y^{2}=-1$. If there is a solution to (\ref{eq:Q(i)_gen}),
then this solution still holds in $\QQ_{p}$. Thus proving that no
solution of (\ref{eq:Qp}) exists would conclude the proof. We first
show that in (\ref{eq:Qp}), $a$ and $b$ are in fact in $\ZZ_{p}$.
In terms of valuation, we have \[
\nu_{p}(a^{2}-pb^{2})=\nu_{p}(y+px).\]
 Since $x\in\ZZ_{p}$ and $y$ is a unit, the right term yields
$\nu_{p}(y+px)\geq\inf\{\nu_{p}(y),\nu_{p}(x)+1\}=0$,
and we have equality since the valuations are distinct. Now the left
term becomes $0=\nu_{p}(a^{2}-pb^{2})=\inf\{2\nu_{p}(a),2\nu_{p}(b)+1\}$.
The only possible case is $\nu_{p}(a)=0$, implying $a\in\ZZ_{p}$
and consequently $b\in\ZZ_{p}$. \\
 We conclude showing that
\begin{equation}\label{eq:norm_modp}
a^{2}-pb^{2}=y+px,~a,b,x,y\in\ZZ_{p}
\end{equation}
 has no solution. Reducing (mod $p\ZZ_{p}$), we see that $y$ has
to be a square in $GF(p)$. Since $y^{2}=-1$, $y^{(p-1)/2}=(-1)^{(p-1)/4}=-1$
by choice of $p\equiv5$~(mod 8). By Proposition \ref{pro:square},
$y$ is not a square, which is a contradiction.
\end{proof}
\begin{remrk}
This result does not hold for $p\equiv1$~(mod 8) since, in this
case, $y^{(p-1)/2}=(-1)^{(p-1)/4}=1$ and we get no contradiction.
The fact that this proof does not work anymore is not enough to restrict
ourselves to the case $p\equiv5$~(mod 8). We thus give a counterexample.
\end{remrk}
\begin{example} \label{ex:Q17}
Consider $K=\QQ\left(\sqrt{17},i\right)$ ,
and $x=\frac{3(i-1)}{4}-\frac{(i-1)\sqrt{17}}{4}$.
It is easy to check that $N_{K/\QQ(i)}(x)=i$.
\end{example}


\subsection{The minimum determinant}

We first show that the ideal $\mathcal{I}$ in (\ref{eq:factor-p})
is principal for all $p\equiv1\textrm{\,(mod\,4)}$. Since $N(\mathcal{I})=p$,
it is enough to show that there exists an element
$\alpha\in \Ic$ with absolute norm $N_{K/\QQ}(\alpha)=p$.
Using the fact that $p=u^{2}+v^{2}$ for some $u,v\in\mathbb{Z}$
(that can be computed)\cite{Samuel}, the element $\alpha=\sqrt{u+iv}$ has
the right norm and generates $\Ic$ (resp. $\overline{\alpha}=\sqrt{u-iv}$
generates $\overline{\mathcal{I}}$). Now, take $\theta=\frac{1+\sqrt{p}}{2}$
and let $\overline{\theta}=\frac{1-\sqrt{p}}{2}$ be its conjugate.
We have $\Oc_{K}=\mathbb{Z}[\theta]$. The codewords have
the form
\[
\mathbf{X}=
\frac{1}{\sqrt{p}}
\left[\begin{array}{cc}
       \alpha(a+b\theta) & \alpha(c+d\theta)\\
       i\bar{\alpha}(c+d\bar{\theta}) & \bar{\alpha}(a+b\bar{\theta})
      \end{array}\right]
\]
with $a,b,c,d\in\mathbb{Z}[i]$.
Each layer of the STBC can be encoded by multiplying the vectors
$(a,b)^T$ and $(c,d)^T$ by the matrix
\[
\left[\begin{array}{cc}
      \alpha & \alpha\theta\\
      \overline{\alpha} & \overline{\alpha\theta}
      \end{array}\right],
\]
which generates the $\ZZ[i]^2$ lattice.
We observe that this lattice generator matrix may require
basis reduction in order to be unitary.

Determinants are given by
\begin{align}
\det(\mathbf{X}) & =\frac{1}{p}N_{K/\QQ(i)}(\alpha)
\left(N_{K/\QQ(i)}(a+b\theta)-iN_{K/\QQ(i)}(c+d\theta)\right).
\label{eq:det22}
\end{align}
As the second term in (\ref{eq:det22}) only takes values in $\ZZ[i]$
and its minimum modulus is equal to $1$ (take for example $a=1$
and $b=c=d=0$), we conclude that
\begin{equation}
\delta_{\min}(\Cc_{\infty})=\frac{1}{p^{2}}|N_{K/\QQ(i)}(\alpha)|^{2}
=\frac{1}{p^{2}}N_{K/\QQ}(\alpha)=\frac{1}{p}~.
\label{eq:min-det-22}
\end{equation}
\begin{remrk}
As $p\equiv5$ (mod 8), the largest minimum determinant is given by
$p=5$ corresponding to the Golden code \cite{Belfiore-2}.
\end{remrk}

%
%

\section{$4\times4$ perfect STBC construction}\label{sec_4}

As for the $2\times2$ case, we consider the transmission of QAM
symbols, thus, the base field is $F = \mathbb{Q}(i)$. Let $ \theta =
\zeta_{15}+\zeta_{15}^{-1}=2\cos(\frac{2\pi}{15})$ and $K$ be
$\QQ(i,\theta)$, the compositum of $\QQ(i)$ and $\QQ(\theta)$. Since
$\varphi(15)=8$ ($\varphi$ is the Euler Totient function),
$[\QQ(\theta):\QQ]=4$, and thus $[\QQ(i,\theta):\QQ(i)]=4$. The
discriminant of $\QQ(\theta)$ is $d_{\QQ(\theta)}=1125$ and the
minimal polynomial $p_\theta(X) = X^4-X^3-4X^2+4X+1$. The extension
$K/\QQ(i)$ is cyclic with generator
$\sigma:\zeta_{15}+\zeta_{15}^{-1}\mapsto\zeta_{15}^{2}+\zeta_{15}^{-2}$.

The corresponding cyclic algebra of degree 4 is $\A=(K/\QQ(i),\sigma,\gamma)$,
that is
\[
\A=1\cdot K\oplus e\cdot K\oplus e^2\cdot K\oplus e^3\cdot K
\]
with $e\in K$ such that $e^{4}=\gamma\in F^*$ and $l e = e\sigma(l)$
for all $l\in K$. In order
to obtain a perfect code, we choose $\gamma=i$.


\subsection{The $\ZZ[i]^{4}$ complex lattice}

We search for a complex rotated lattice $\ZZ[i]^4$ following the
approach given in \ref{sub:Complex-algebraic-lattices}.
Since the relative discriminant of $K$ is $d_{K/F}=d_{\QQ(\theta)}=1125=3^{2}\cdot5^{3}$,
a necessary condition to obtain a rotated version of $\ZZ[i]^4$ is that there
exists an ideal $\Ic\subseteq\Oc_{K}$ with norm $45=3^{2}\cdot5$.
The geometrical intuition is that the sublattice $\Lambda(\mathcal{I})$
has fundamental volume equals to
$2^{-4}\sqrt{d_K}N(\Ic)=3^{4}\cdot5^{4}=\sqrt{15}^{8}$,
which suggests that the fundamental parallelotope of the algebraic
lattice $\Lambda(\Ic)$ could be a hypercube of edge length
equal to $\sqrt{15}$.

An ideal $\Ic$ of norm $45$ can be found from the following
ideal factorizations

\begin{eqnarray*}
(3)\Oc_{K} & = & \Ic_{3}^{2}\overline{\Ic_{3}}^{2}\\
(5)\Oc_{K} & = & \Ic_{5}^{4}\overline{\Ic_{5}}^{4}
\end{eqnarray*}

Let us consider $\Ic=\Ic_{3}\cdot\Ic_{5}$.
It is a principal ideal $\Ic=(\alpha)$ generated
by $\alpha=(1-3i)+i\theta^{2}$.

A $\ZZ[i]$--basis of $(\alpha)$ is given by $\{\alpha\theta^{i}\}_{i=0}^{3}$.
Using the change of basis given by the following matrix \[
\left(\begin{array}{cccc}
1 & 0 & 0 & 0\\
0 & 1 & 0 & 0\\
0 & -3 & 0 & 1\\
-1 & -3 & 1 & 1\end{array}\right),\]
 one gets a new $\ZZ[i]$--basis
\[
\{\nu_{k}\}_{k=1}^{4}=\left\{(1-3i)+i\theta^{2},(1-3i)\theta+i\theta^{3},
-i+(-3+4i)\theta+(1-i)\theta^{3},(-1+i)-3\theta+\theta^{2}+\theta^{3}\right\}.
\]
Then by straightforward computation we can check that
\[
\frac{1}{15}\Tr_{K/\QQ(i)}(\nu_{k}\bar{\nu_{\ell}})=\delta_{k\ell}~~~~~k,\ell=1,\ldots,4
\]
 using $
\Tr_{\QQ(\theta)/\QQ}(\theta)=1,~\,
\Tr_{\QQ(\theta)/\QQ}(\theta^{2})=9,\,\,
\Tr_{\QQ(\theta)/\QQ}(\theta^{3})=1,~\,
\Tr_{\QQ(\theta)/\QQ}(\theta^{4})=29$.
For example, we compute the diagonal coefficients,
\[
\Tr_{K/\QQ(i)}(|\nu_{k}|^{2})
=\left\{ \begin{array}{cc}
\Tr_{K/\QQ(i)}(10-6\theta^{2}+\theta^{4})=15 & \mbox{if}~~k=1\\
\Tr_{K/\QQ(i)}(1+3\theta+\theta^{2}-\theta^{3})=15 & \mbox{if}~~k=2\\
\Tr_{K/\QQ(i)}(5+6\theta-\theta^{2}-2\theta^{3})=15 & \mbox{if}~~k=3\\
\Tr_{K/\QQ(i)}(-5\theta+2\theta^{2}+2\theta^{3})=15 & \mbox{if}~~k=4
\end{array}\right..
\]
The unitary generator matrix of the lattice is given by
\begin{align*}
\Rm & = \frac{1}{\sqrt{15}} (\sigma_{\ell}(\nu_{k}))_{k,\ell=1}^{n}\\
 & =\left(\begin{array}{cccc}
0.2582-0.3122i & 0.3455-0.4178i & -0.4178+0.5051i & -0.2136+0.2582i\\
0.2582+0.0873i & 0.4718+0.1596i & 0.1596+0.054i & 0.7633+0.2582i\\
0.2582+0.2136i & -0.5051-0.4178i & -0.4178-0.3455i & 0.3122+0.2582i\\
0.2582-0.7633i & -0.054+0.1596i & 0.1596-0.4718i & -0.0873+0.2582i
\end{array}\right).
\end{align*}


\subsection{The norm condition}

We now show that $\A=(K/\QQ(i),\sigma,i)$ is a division algebra.
By Proposition \ref{pro:norm}, we have to check that $\pm i$ and
$-1$ are not norms of elements in $K$.

\begin{lemma}
\label{lem:Q5}We have the following field extensions:\[
\mathbb{Q}(i)\subset\mathbb{Q}(i,\sqrt{5})\subset K\]

\end{lemma}
\begin{proof}
We show that $\QQ(i,\sqrt{5})$ is the subfield fixed by $\langle\sigma^{2}\rangle$,
the subgroup of order 2 of Gal$(K/\QQ(i))=\langle\sigma\rangle$.
Let
$\sigma^{2}:\zeta_{15}+\zeta_{15}^{-1}\mapsto\zeta_{15}^{4}+\zeta_{15}^{-4}$
and $x=\sum_{k=0}^{3}a_{k}(\zeta+\zeta^{-1})^{k}$, $a_{k}\in\QQ(i)$,
be an element of $K$. It is a straightforward computation to show
that $\sigma^{2}(x)=x$ implies that $x$ is of the form
$x=a_0+a_3(\zeta_{15}^{3}+\zeta_{15}^{-3})=a_0+a_3\dfrac{-1+\sqrt{5}}{2}\in \QQ (i,\sqrt{5})$.
\end{proof}
\begin{prop}
The algebra $\A=(K/\QQ(i),\sigma,i)$ is a division algebra.
\end{prop}
\begin{proof}
We start by proving by contradiction that $\pm i$ are not a norm.
Suppose $\pm i$ is a norm in $K^{*}$, i.e., there exists $x\in K^{*}$
such that $N_{K/\mathbb{Q}(i)}(x)=\pm i$. By Lemma \ref{lem:Q5}
and transitivity of the norm, we have
\[
N_{K/\QQ(i)}(x)=N_{\QQ(i,\sqrt{5})/\QQ(i)}(N_{K/\QQ(i,\sqrt{5})}(x))=\pm i.
\]
 Thus $\pm i$ has to be a norm in $\QQ(i,\sqrt{5})$.
By Proposition \ref{pro:not-norm} in the case $p=5$, we know $i$ is not
a norm. In order to show that $-i$ is not a norm, it is enough to slightly
modify the proof of Proposition \ref{pro:not-norm}. Eq.~(\ref{eq:norm_modp})
becomes, with $p=5$,
\[
a^2 - 5b^2 = -y + 5x,~a,b,x,y \in \ZZ_5.
\]
Reducing (mod 5), we see that in order for this equation to have a solution,
$y$ has to be square in $GF(5)$.
Since $(-y)^{(p-1)/2}=(-y)^2=y^2=-1$, $y$ cannot be a square
(see Proposition \ref{pro:square}) and we get
a contradiction.

The previous argument does not apply for $-1$ since it is clearly a norm
in $\QQ(i,\sqrt{5})/\QQ(i)$. The proof that $-1$ is not a
norm uses techniques from Class field theory
and is given in  Appendix \ref{sec:-1-not-norm-4}.
\end{proof}


\subsection{The minimum determinant}

From (\ref{eq:min-det-principal}), or similarly from Proposition \ref{prop:dmin_principal},
the minimum determinant of the infinite code is equal to
\[
\delta_{min}(\mathcal{C}_{\infty})
=\frac{1}{15^{4}}\cdot N_{K/\QQ}(\alpha)
=\frac{45}{15^4}
=\frac{1}{1125}=\frac{1}{d_{\QQ(\theta)}}.
\]

%
%

\section{$3\times3$ perfect STBC construction}\label{sec_3}

In this case we use HEX symbols. Thus, the base field is $F=\QQ(j)$.
Let $\theta = \zeta_{7}+\zeta_{7}^{-1}=2\cos(\frac{2\pi}{7})$ and
$K$ be $\QQ(j,\theta )$, the compositum of $F$ and $\QQ(\theta)$.
Since $\varphi(7)=6$, $[\QQ(\theta):\QQ]=3$, and thus
$[\QQ(j,\theta):F]=3$. The discriminant of $\QQ(\theta)$ is
$d_{\QQ(\theta)}=49$ the minimal polynomial $p_\theta(X) =
X^3+X^2-2X-1$. The extension $K/F$ is cyclic with generator
$\sigma:\zeta_{7}+\zeta_{7}^{-1}\mapsto\zeta_{7}^{2}+\zeta_{7}^{-2}$.

The corresponding cyclic algebra of degree 3 is $\A=(K/F,\sigma,\gamma)$,
that is
\[
\A=1\cdot K\oplus e\cdot K\oplus e^2\cdot K
\]
with $e\in\A$ such that $e^{3}=\gamma\in F^*$ and $l e = e\sigma(l)$
for all $l\in K$.
In order to obtain a perfect code, we choose $\gamma=j$.


\subsection{The $\ZZ[j]$--lattice  $\ZZ[j]^{3}$ }

In this case, we look for a $\ZZ[j]$-lattice which is a rotated
$\ZZ[j]^3(=A_2^3)$ lattice. The relative discriminant of $K$ is
$d_{K/F}=d_{\QQ(\theta)}=49=7^{2}$, while its absolute discriminant
is $d_{K}=-3^{3}\cdot7^{4}$. A necessary condition to obtain a
rotated $\ZZ[j]^{3}$ lattice is the existence of an ideal
$\Ic\subseteq\Oc_{K}$ with norm $7$. In fact, the lattice
$\Lambda(\Oc_{K})$ has fundamental volume equal to
$2^{-3}\sqrt{|d_{K}|}=7^{2}\left(\frac{\sqrt{3}}{2}\right)^{3}$ and
the sublattice $\Lambda(\Ic)$ has fundamental volume equals to
$2^{-3}\sqrt{|d_{K}|}N(\Ic)=7^{3}\left(\frac{\sqrt{3}}{2}\right)^{3}$,
where the norm of the ideal $N(\Ic)$ is equal to the sublattice
index. This suggests that the algebraic lattice $\Lambda(\Ic)$ could
be a homothetic (scaled rotated) version of $A_{2}^{3}$, namely,
$\left(7A_{2}\right)^{3}$.

An ideal $\mathcal{I}$ of norm $7$ can be found from the following
ideal factorizations
\begin{eqnarray*}
(7)\Oc_{K} & = & \Ic_{7}^3\overline{\Ic_{7}}^3.
\end{eqnarray*}
Let us consider $\Ic=\Ic_{7}$. It is a principal
ideal $\Ic=(\alpha)$ generated by $\alpha=(1+j)+\theta$.
A $\ZZ[j]$--basis of $(\alpha) {\cal O}_K$ is given by
$\{\alpha\theta^{k}\}_{k=0}^{2}
=\{(1+j)+\theta,(1+j)\theta+\theta^{2},
1+2\theta+j\theta^{2}\}$.
Using the change of basis given by the following matrix
\[
\left(\begin{array}{ccc}
1 &  0 & 0\\
0 & -1 & 1\\
2 &  1 & 0\end{array}\right),\]
 one gets a reduced $\ZZ[j]$--basis
\[
\{\nu_{k}\}_{k=1}^{3}
=\{(1+j)+\theta,(-1-2j)+j\theta^{2},
(-1-2j)+(1+j)\theta+(1+j)\theta^{2}\}.
\]
Then by straightforward computation we find
\[
\frac{1}{7}\Tr_{K/\QQ(j)}(\nu_k\bar{\nu_l})=
\delta_{kl}~~~~~k,l=1,2,3
\]
using $\Tr_{\QQ(\theta)/\QQ}(1)=3,~\Tr_{\QQ(\theta)/\QQ}(\theta)=-1,
~\Tr_{\QQ(\theta)/\QQ}(\theta^{2})=5.$

We compute, for example, the diagonal coefficients
\[
\Tr_{K/\QQ(j)}(\nu_k\bar{\nu_k})=\left\{ \begin{array}{cc}
\Tr_{K/\QQ(j)}(1+\theta+\theta^{2})=7 & \mbox{if}~~k=1\\
\Tr_{K/\QQ(j)}(2-\theta)=7 & \mbox{if}~~k=2\\
\Tr_{K/\QQ(j)}(4-\theta^{2}) =7& \mbox{if}~~k=3\end{array}\right.
\]
The generator matrix of the lattice in its numerical form is thus given by
\begin{align*}
\Rm & =\dfrac{1}{\sqrt{7}}(\sigma_l(\nu_k))_{k,l=1}^{n}\\
  & = \left(\begin{array}{ccc}
     0.66030 + 0.32733i &  0.02077 + 0.32733i & -0.49209 + 0.32733i\\
    -0.29386 - 0.14567i & -0.03743 - 0.58982i & -0.61362 + 0.40817i\\
     0.52952 + 0.26250i & -0.04667 - 0.73550i &  0.27309 - 0.18165i
     \end{array}\right).
\end{align*}


\subsection{The norm condition}

We show that the rank criterion is fullfilled by this new code. The
following proposition guarantees that
$\A=(\QQ(j,\theta)/F,\sigma,j)$ is a division algebra.

\begin{prop}
The units $j$ and $j^{2}$ are not norms in
$\mathbb{Q}(j,\theta)/F$.
\end{prop}
\begin{proof}
See Appendix \ref{sec:j-not-norm-3} for the proof,
which uses Class Field Theory.
\end{proof}

\subsection{The minimum determinant}

As the ideal $\Ic$ is principal, we can use
(\ref{eq:min-det-principal}) or Proposition \ref{prop:dmin_principal} to get
\[
\delta_{min}(\Cc_{\infty})
=\frac{1}{7^{3}} N_{K/\QQ}(\alpha)
=\frac{7}{7^3}
=\frac{1}{49}=\frac{1}{d_{\QQ(\theta)}}.
\]

%
%

\section{$6\times6$ perfect STBC construction}\label{sec_6}

As in the $3$ antennas case, we transmit HEX symbols. Thus,
the base field is $F=\QQ(j)$.
Let $\theta = \zeta_{28}+\zeta_{28}^{-1}=2\cos(\frac{\pi}{14})$ and $K$ be
$\QQ(j,\theta)$, the compositum of $F$ and $\QQ(\theta)$.
Since $\varphi(28)=12$, $[\QQ(\theta):\QQ]=6$,
and thus $[\QQ(j,\theta):F]=6$.
The extension $K/F$ is cyclic with generator
$\sigma:\zeta_{28}+\zeta_{28}^{-1}\mapsto\zeta_{28}^{2}+\zeta_{28}^{-2}$.

The corresponding cyclic algebra of degree 6 is $\A=(K/F,\sigma,\gamma)$,
that is
\[
\A=1\cdot K \oplus e\cdot K\oplus e^2\cdot K \oplus e^3 \cdot K
    \oplus e^4 \cdot K \oplus e^5 \cdot K
\]
with $e\in\A$ such that $e^{6}=\gamma\in F^*$ and $l e = e\sigma(l)$
for all $l\in K$.
In order to obtain a perfect code, we choose $\gamma=-j$.


\subsection{The $\ZZ[j]$-lattice $\mathbb{Z}[j]^6$}

First note that the discriminant of $K$ is $d_K=2^{12}\cdot3^{6}\cdot7^{10}$.
Following the approach given in Section \ref{sub:Complex-algebraic-lattices},
we need to construct a $\ZZ[j]^6$ lattice.

A necessary condition to obtain a rotated version of
$\mathbb{Z}[j]^{6}$ is that there
exists an ideal $\mathcal{I}\subseteq\Oc_{K}$ with norm $7$.
In fact, the lattice $\Lambda(\Oc_{K})$ has fundamental volume equal to
$2^{-6}\sqrt{|d_{K}|}=7^{5}\cdot2^{6}\cdot\left(\frac{\sqrt{3}}{2}\right)^{6}$
and the sublattice $\Lambda(\Ic)$ has fundamental volume equal to
$2^{-6}\sqrt{|d_K|}N(\Ic)=7^{6}\cdot2^6\cdot\left(\frac{\sqrt{3}}{2}\right)^6$,
where the norm of the ideal $N(\Ic)$ is equal to the sublattice
index. This suggests that the algebraic lattice $\Lambda(\Ic)$
could be a homothetic version of $A_{2}^{6}$, namely,
$\left(\sqrt{14}A_{2}\right)^{3}$, but this needs to be checked explicitly.

An ideal $\Ic$ of norm $7$ can be found from the following
ideal factorizations
\begin{eqnarray*}
(7)\Oc_{K} & = & \Ic_{7}^6\overline{\Ic_{7}}^6.
\end{eqnarray*}
Let us consider $\Ic=\Ic_7$. Unlike in the preceeding constructions,
the ideal $\Ic$ is not principal. This makes harder the explicit computation
of an ideal basis, and in particular of the ideal basis (if any) for which the
Gram matrix becomes the identity.

We thus adopt the following alternative approach.
We compute numerically a basis of $\Ic$, from which we compute a Gram
matrix of the lattice. We then perform a basis reduction on the Gram
matrix, using an LLL reduction algorithm (see Appendix \ref{sec:LLL} for
more details). This gives both the Gram matrix in the reduced basis
and the matrix of change of basis. We get the following
change of basis
\[
\left(\begin{array}{cccccc}
  0   & 1 & 0 & 0 & 0 & 0 \\
1+j   & 0 & 1 & 0 & 0 & 0 \\
-1-2j & 0 & -5& 0 & 1 & 0\\
 1+j  & 0 & 4 & 0 & -1& 0 \\
  0   & -3& 0 & 1 & 0 & 0\\
  0   & 5 & 0 & -5& 0 & 1
\end{array}\right)
\]
and the lattice
generator matrix in numerical form
{\small
\[
\Rm=\frac{1}{\sqrt{14}} \left(\begin{array}{cccccc}
 1.9498         &  1.3019-0.8660i & -0.0549-0.8660i &
-1.7469-0.8660i &  1.5636         &  0.8677  \\
 0.8677         & -1.7469-0.8660i &  1.3019-0.8660i &
-0.0549-0.8660i & -1.9498         &  1.5636 \\
 1.5636         & -0.0549-0.8660i & -1.7469-0.8660i &
 1.3019-0.8660i & -0.8677         & -1.9498 \\
-1.9498         &  1.3019-0.8660i & -0.0549-0.8660i &
-1.7469-0.8660i & -1.5636         & -0.8677 \\
-0.8677         & -1.7469-0.8660i &  1.3019-0.8660i &
-0.0549-0.8660i &  1.9498         & -1.5636 \\
-1.5636         & -0.0549-0.8660i & -1.7469-0.8660i &
 1.3019-0.8660i &  0.8677         &  1.9498\\
\end{array}\right).
\]
}
This matrix satisfies $\Rm \Rm^H$ is the identity matrix, so that we indeed
get a rotated version of the $A_2^6$ lattice.


\subsection{The norm condition}

Since $\gamma=-j$, we have to check that $-j,j^2,-j^3=-1,j^4=j$ and
$-j^5=-j^2$ are not norms in $K$.

\begin{lemma}\label{lem:Qcos7}
We have the following field extensions:
\[
\QQ(j) \subset \QQ(j,\zeta_7+\zeta_7^{-1})
       \subset \QQ(j,\zeta_{28}+\zeta_{28}^{-1})
\]
\end{lemma}
\begin{proof}
The proof is similar to that of Lemma \ref{lem:Q5}.
One has to show that $\QQ(j,\zeta_7+\zeta_7^{-1})$ is the subfield
fixed by $\langle\sigma^2\rangle$, the subgroup of order 2 of
$\Gal(\QQ(j,\zeta_{28}+\zeta_{28}^{-1})/F)=\langle\sigma\rangle$.
\end{proof}
\begin{prop}
The algebra $\A=(K/F,\sigma,-j)$ is a division algebra.
\end{prop}
\begin{proof}
We prove, by contradiction, that $\pm j$ and $\pm j^2$ are not norms
in $K^*$. Suppose that either $\pm j$ or $\pm j^2$ are a norm in $K^*$, i.e.,
there exists $x \in K^*$ such that $N_{K/F}(x)=\pm j$ (resp. $\pm j^2$).
By Lemma \ref{lem:Qcos7} and transitivity of the norm,
we have
\begin{equation}\label{eq:normQcos28}
N_{K/F}(x)=N_{\QQ(j,\zeta_7+\zeta_7^{-1})/F}
        (N_{K/\QQ(j,\zeta_7+\zeta_7^{-1})}(x))=\pm j ~~~(\mbox{resp.}~\pm j^2).
\end{equation}
Thus $j$ and $j^2$ have to be a norm in $\QQ(j,\zeta_7+\zeta_7^{-1})$,
which is not the case, by Propositions \ref{prop:j-not-norm} and
\ref{prop:j2-not-norm} in Appendix \ref{sec:j-not-norm-3}.\\
For the cases of $-j$ and $-j^2$, since $[\QQ(j,\zeta_7+\zeta_7^{-1}):F]=3$,
(\ref{eq:normQcos28}) yields
\[
N_{\QQ(j,\zeta_7+\zeta_7^{-1})/F}
 (-N_{K/\QQ(j,\zeta_7+\zeta_7^{-1})}(x))= j ~~~(\mbox{resp.}~j^2),
\]
which gives the same contradiction.

The proof that -1 is not a norm can be found in Appendix
\ref{sec:-1-not-norm-6} and uses Class Field Theory.
\end{proof}


\subsection{The minimum determinant}

Since the ideal $\Ic$ is not principal,
we use the bounds of Corollary \ref{cor:bound_dmin}
\[
\frac{1}{14^{6}}
\cdot N_{K/\QQ}(\Ic) =
\frac{1}{2^{6}\cdot7^{5}}=\frac{1}{d_{\QQ(\theta)}}
\leq
\delta_{min}(\mathcal{C}_{\infty})
\leq
\frac{1}{14^6}\min_{x\in \Ic}N(x)= \frac{7^2}{2^6 \cdot 7^6}
\]
yielding
\[
\dfrac{1}{2^{6}\cdot7^{5}} \leq \delta_{min}(\mathcal{C}_{\infty}) \leq
\dfrac{1}{2^{6}\cdot7^{4}}.
\]

%
%

\section{Existence of perfect codes}\label{sec:no_other}

Since we have given constructions only for dimensions  $2$, $3$, $4$
and $6$, it is interesting to discuss the existence of perfect
codes. Perfect space-time block codes must satisfy a large number of
constraints. Let us derive here the consequences of these
constraints in the choice of the corresponding cyclic algebra.

First note that in order to have non vanishing determinants when the
spectral efficiency increases, determinants of the infinite code
$\Cc_{\infty}$ must take values in a discrete subset of $\CC$.
We have shown in Section \ref{subsec:discrete_det} that
the determinants of $\Cc_{\Ic}$ are in $\Oc_{F}$,
when $\Ic \subseteq \Oc_K$ and $\gamma \in \Oc_F$.
But $\Oc_{F}$ is discrete in $\mathbb{C}$ if and only if $F$ is a
quadratic imaginary field, namely $F=\QQ(\sqrt{-d})$,
with $d$ a positive square free integer. Indeed, we have that
$|a+b\sqrt{-d}|^2 \in \ZZ$ if $a,b \in \ZZ$. The positive minimum of
an integer is thus 1. This is not true anymore if we consider
already $|a+b\sqrt{d}|^2$, which belongs to $\ZZ[\sqrt{d}]$. We cannot
obtain a minimum without any constraint on $a,b \in \ZZ$. The same
phenomenon appears even more clearly in higher dimension.

The average energy per antenna constraint requires $|\gamma|=1$.
Furthermore, the proof of the non-vanishing determinant relied on
$\gamma$ being in $\Oc_F$. There are two ways of getting a tradeoff
between these two conditions. Our approach consists in choosing
$\gamma$ to be a root of unity. Since the base field has to be
quadratic, this gives as choice $\QQ(i)$, which contains the 4th
root of unity $i$, and $\QQ(j)$, which contains the 3rd root of
unity $j$ and the 6th root of unity $-j$. The following lemma
confirms these are the only possibilites:
\begin{lemma}\label{lem:quadratic-units}\cite[p.76]{Samuel}
Let $d$ be a positive square free integer.
The only units of $F=\QQ(\sqrt{-d})$
are $\pm1$ unless $F=\QQ(i)$ or $F=\QQ(j)$.
\end{lemma}
As a consequence, the perfect codes proposed are available only
in dimension $2$, $3$, $4$ and $6$.

Elia et al. recently considered the option of droping one
of the two conditions.  In \cite{Elia}, they drop the constraint
$|\gamma|=1$, at the price of loosing the average energy advantage.
They also consider an element $\gamma$ of norm 1, but not in $\Oc_F$.
Since $\gamma=\gamma_1/\gamma_2 \in F$, the minimum determinant of
the resulting code can be written as
$\frac{1}{|\gamma_2|^{2(n-1)}}\det(\tilde{\Xm})$, where $\tilde{\Xm}$
is a codeword with coefficients in $\Oc_K$. Thus the non-vanishing
determinant property holds, but there is a loss in the coding gain
proportional to $|\gamma_2|^{2(n-1)}$. These codes are not
restricted to the dimensions $2$, $3$, $4$ and $6$.

%
%
\section{Simulation results}

We have simulated the complete MIMO transmission scheme using perfect
Space-Time codes, and the previously best known codes. Transmitted
symbols belong to $q$-QAM ($2$ and $4$ antennas) or $q$-HEX ($3$
antennas) constellations, $q=4,8,16,64$. We used the modified
version of the Sphere-decoder presented in \cite{Rekaya-2}.

QAM constellations have minimum Euclidean distance $2$. The
respective average energy per symbol for the $4,8,16\mbox{ and
}64$-QAM constellations are $2,6,10$ and $42$. The $q$-HEX
constellations are finite subsets of the hexagonal lattice $A_{2}$.
In fact the hexagonal lattice is the densest lattice in dimension
$2$; constellations using points from the hexagonal lattice ought to
be the most efficient \cite{Forney-3}. Since $A_{2}$ is not a binary
lattice, bit labeling and constellation shaping must be performed
{\em ad hoc}. The best finite hexagonal packings for the desired
sizes are presented in Figures \ref{cap:Hex-const-4},
\ref{cap:Hex-const-8}, and \ref{cap:Hex-const-16}.

The respective average energy per symbol for the $4,8,$ and $16$ HEX
constellations with minimum Euclidean distance $2$ are $2$, $4.5$
and $8.75$. We should note the energy saving compared to QAMs of the
same size. The HEX constellations are carved from shifted versions
of the lattice $2A_2$. For $4,8$-HEX constellations the shift is in
$(1,0)$, while for $16$-HEX constellation the shift is in $(1/2,0)$.

In Figure \ref{cap:fig22}, we have plotted the codeword error rates
for the Golden code (GC), some other $2\times2$ Perfect codes (PC)
and the best previously known $2\times2$ STBCs \cite{Damen-3} (BPC),
as a function of $E_{b}/N_{0}$, using $4, 16, 64$-QAM
constellations. In \cite{Damen-3}, the values of $\gamma$ giving the
best codes were obtained by numerical optimizations and depend on
the spectral efficiency. As we concluded in
\cite{Belfiore-3,Belfiore-2}, the Golden code has the best
performance. We see in Fig. \ref{cap:fig22} that perfect codes with
$p=\sqrt{13}$ and $p=\sqrt{37}$ have performance close to that of
the BPCs. However the code with $p=\sqrt{17}$ which is not a perfect
code (the cyclic algebra is not a division algebra) has the worst
performances, and we can even observe a change in the slope of the
curve for high SNR, due to the reduced diversity order of this code
($2$ instead of $4$). In fact, as shown in Example \ref{ex:Q17},
there exists an $x\in K=\mathbb{Q}(i,\sqrt{17})$ such that
$N_{K/\mathbb{Q}(i)}(x)=i$. The appearance of such an $x$ is rare,
which explains why this code works well at low and medium signal to
noise ratio and the change of slope appears at very low error rates.

In Fig. \ref{cap:fig33} and \ref{cap:fig44}, we have plotted
respectively the codeword error rates of the $3\times3$ and the
$4\times4$ PC and the best previously known codes \cite{El-Gamal-2,
Galliou-2} as a function of $E_{b}/N_{0}$. In Fig. \ref{cap:fig33},
we see that for the 4-HEX constellation, the BPC performs  a little
better than the PC. However, when the constellation is 8-HEX or
larger, PCs have better performance, due to the constant minimum
determinant.

In Fig. \ref{cap:fig44}, we note that the $4\times 4$ PC improves
over BPC codes when we use the 64-QAM.

%
%

\section{Conclusion}\label{sec:conclusion}

In this paper we presented new algebraic constructions of full-rate,
fully diverse $2\times2$, $3\times3$, $4\times4$ and $6\times6$ space-time
codes, having a constant minimum determinant as the spectral efficiency
increases. The name {\em perfect STBC}, used for  these codes, was
suggested by the fact that they satisfy a large number of design criteria
and only appear in a few special cases as the
classical perfect error correcting codes,
achieving the Hamming sphere packing bound.

\appendices{~}

%
%

\section{Number Fields: Basic Definitions}\label{app:basics}

The codebooks we build are based on cyclic algebras built over number
fields, we thus need some background on number fields. This appendix
aims at giving intuition to the reader who does not know the topic.
It focuses on examples, and may skip some technical points in
order to be more accessible.

Number fields can first be thought of as finite vector spaces over a
base field. For example, $\QQ(i)=\{a+bi,~a,b \in \QQ\}$ is a vector
space of dimension 2 over $\QQ$, whose basis is given by $\{1,i\}$.
In our case, we will consider two number fields, denoted by $K$ and
$F$, and $K$ will be a vector space of dimension $n$ over $F$. We say
that $K$ is a field {\em extension} of $F$, which we denote by $K/F$.
The dimension of $K$ over $F$ as a vector space is called the
{\em degree}, and is denoted by $[K:F]$.
Another way of thinking of a number field is to add a root of a
polynomial, with coefficients in $F$, to
a field, and to add also all its powers and multiples, so that the
resulting set is indeed a field. For example, $\QQ(i)$ is built adding
the roots of the polynomial $X^2+1$ to $\QQ$. The field extension
$K/F$ can similarly be seen as adding the element $\theta$, root of a
polynomial $p(X)$, to $K$. We may write $K=F(\theta)$. Since a
polynomial has $n$ roots, one may wonder if taking one root or another
may change the number field. If all the roots are indeed in the number
field, it does not change, and the number field is called a
{\em Galois extension}. Not all number fields are Galois extensions.

For our purpose, we are interested in a field extension $K/F$
such that all roots $\theta_1,\ldots,\theta_n$ of $p(X)$ are not only
in $K$, but furthermore are related to each other as follows: there
exists a map $\sigma$ such that $\sigma^k(\theta_1)=\theta_j$,
$k,j=1,\ldots,n$. In such case, $K/F$ is called a {\em cyclic Galois}
extension, and $\{\sigma^k\}$, $k=1,\ldots,n$, is called a
{\em (cyclic) Galois group} (it can be shown that it has indeed a group
structure). For example, $\QQ(i)$ is a cyclic Galois extension of degree 2,
since there exists $\sigma: i \mapsto -i$.

There are two important objects that can be defined thanks to
$\{\sigma^k\}$, $k=1,\ldots,n$. We define the {\em trace} and
the {\em norm} of an element $x \in K$ resp. as follows:
\[
\mbox{Tr}_{K/F}(x)=\sum_{k=0}^{n-1}\sigma^k(x),~
\mbox{N}_{K/F}(x)=\prod_{k=0}^{n-1}\sigma^k(x).
\]
We may call a relative trace/norm if the base field is not $\QQ$, by
opposition to an absolute trace/norm when $F=\QQ$.

Let now $L$ be a number field of degree $n$ over $\QQ$.
Consider the set of elements $x$ of $L$  that satisfy the following
property: there exists a monic polynomial $f$ with
coefficients in $\ZZ$ such that $f(x)=0$. This set is called the
{\em  ring of integers} of $L$, and is denoted by $\Oc_L$. It can be
shown that this is indeed a ring, but what is more interesting is that
this set posseses a $\ZZ$-basis. We will use this fact extensively in
the paper. Let $\{\omega_1,\ldots,\omega_n \}$ be a $\ZZ$-basis, i.e.,
all elements can be written as integer linear combinations of basis elements.
Then $\det(\mbox{Tr}_{L/\QQ}(\omega_i\omega_j)_{i,j=1}^n)$ is an invariant of
$L$ called the {\em discriminant} of $L$. Similarly as for the trace
and norm, we call the discriminant {\em absolute} to emphasize that
the base field is $\QQ$, and $relative$ otherwise.

%
%

\section{The Hasse Norm Symbol}\label{sec:Hasse}

In this Appendix, we introduce {\em the Hasse Norm Symbol}. It is a
tool derived from Class Field Theory, that allows to compute whether a
given element is a norm. Our exposition is based on \cite{Gras}.
In the following, we consider extensions of number fields $K/F$ that we
assume abelian.

Denote by $K_{\nu}$ the completion of $K$ with respect
to the valuation $\nu$. We denote the {\em embedding} of $K$ into
$K_{\nu}$ by $i_{\nu}$.
\begin{definitn}\label{def:Hasse}\cite[p.~105]{Gras}
Let $K/F$ be an abelian extension of number fields with Galois group
$\Gal(K/F)$.
The map
\[
\begin{array}{clll}
\left(\frac{\bullet~,~K/F}{\nu}\right):  & K^* & \rightarrow & \Gal(K/F) \\
             & x   & \mapsto     & \left(\frac{i_{\nu}(x),K/F}{\nu}\right)
\end{array}
\]
is called the {\em Hasse norm symbol}.
\end{definitn}
The main property of this symbol is that it gives a way to compute
whether an element is a local norm \cite[p.~106,107]{Gras}.
\begin{thm}
We have $\left(\frac{x,K/F}{\nu}\right)=1$ if and only if $x$ is a
local norm at $\nu$ for $K/F$.
\end{thm}
In order to compute the Hasse norm symbol, we need to know some of
its properties. Let us begin with a property of linearity.
\begin{thm}
We have
\[
\left(\frac{xy,K/F}{\nu}\right)=
\left(\frac{x,K/F}{\nu}\right)\left(\frac{y,K/F}{\nu}\right).
\]
\end{thm}
We then know how the symbol behaves at
unramified places \cite[p.~106]{Gras}.
\begin{thm}\label{th:unramified}
If $\nu$ is unramified in $K/F$, then we have, for all $x \in F^*$:
\[
\left(\frac{x,K/F}{\nu}\right) = \left(\frac{K/F}{\nu}\right)^{v(x)},
\]\
where $\left(\frac{K/F}{\nu}\right)$ denotes the Frobenius of $\nu$
for $K/F$ (see Remark \ref{rem:Frob} below), and $v(x)$
denotes the valuation of $x$.
\end{thm}
\begin{remrk}\label{rem:Frob}
For our purpose, it is enough to know that the Frobenius
$\left(\frac{K/F}{\nu}\right)$ is an element of the Galois group
$\Gal(K/F)$. We do not need to know it explicitly. For a precise
definition, we let the reader refer to \cite[p.~107]{Gras}.
\end{remrk}
\begin{cor}\label{cor:unit_norm}
At an unramified place, a unit is always a norm.
\end{cor}
\begin{proof}
It is straightforward since the valuation of a unit is 0.
\end{proof}
A remarkable property of the Hasse norm symbol is the {\em product
formula} \cite[p.~113]{Gras}.
\begin{thm}\label{th:product_formula}
Let $K/F$ be a finite extension. For any $x \in F^*$ we have:
\[
\prod_{\nu}\left(\frac{x,K/F}{\nu}\right)=1,
\]
where the product is defined over all places $\nu$.
\end{thm}
\begin{remrk}
By Corollary \ref{cor:unit_norm}, we know that a unit is always a
norm locally if the place is unramified. Since we are interested in
showing that a unit $\gamma$ is not a norm, we will look for a contradiction
at a ramified place.
\end{remrk}

Before giving the proofs in themselves, we explain briefly their general
scheme.
The idea is to start from the product formula, and to simplify all the
terms except two in the product over all primes, so that we get
a product of two terms equal to 1:
\[
\left(\frac{\gamma,K/F}{\nu}\right)\left(\frac{x,K/F}{\nu'}\right)=1,
~x \in F^*.
\]
Hopefully, one of the two terms left will involve $\gamma$,
the other will be shown to be different from 1, so that since the product
is 1, we will deduce that the term involving $\gamma$ is different from 1,
thus $\gamma$ is not a norm.
In order to make it easier to simplify the product formula,
we introduce an element $y \in K$ such that $y\gamma$ is a unit locally
at ramified primes, and we compute the product formula
\[
\prod_{\nu}\left(\frac{y\gamma,K/F}{\nu}\right)=1.
\]

%
%

\section{$j$ and $j^2$ are not a norm~ in
$\QQ(j,2\cos\left(\frac{2\pi}{7}\right))/\QQ(j)$\label{sec:j-not-norm-3}}

In this section, we prove that $j$ and $j^2$ are a not a norm in
$\QQ(j,2\cos\left(\frac{2\pi}{7}\right))/\QQ(j)$. We show that $j$
and $j^2$ are not a norm locally by computing their Hasse norm
symbol. The proof is detailed for $j$.
\begin{prop}\label{prop:j-not-norm}
The unit $j$ is not a norm in
$K/F=\QQ(j,2\cos\left(\frac{2\pi}{7}\right))/\QQ(j)$.
\end{prop}
\begin{proof}
We consider the field extension $K/F$. We have
\[
7\ZZ[j]=(j-2)(j+3)=\mathfrak{p}_{7}\mathfrak{q}_{7}.
\]
We show that $j$ is not a norm locally in $\mathfrak{p}_{7}$,
thus $j$ is not a norm in $K$.\\

We look for a number $y$ in $\ZZ[j]$ satisfying
\begin{eqnarray}
y  & \equiv & 1~(\mbox{mod }j-2)\label{y_cond1}\\
jy & \equiv & 1~(\mbox{mod }j+3)\label{y_cond2}.
\end{eqnarray}
By applying the Chinese Remainder Theorem over $\ZZ[j]$, we find $y=7-3j$ with
$(y)\ZZ[j]=\mathfrak{p}_{79}$. Let $\left(\frac{x,K/F}{\nu}\right)$
denote the Hasse norm symbol. By the product formula
\begin{equation}
\prod_{\nu}\left(\frac{jy,K/F}{\nu}\right)
=\prod_{\nu\mbox{ ramified}}\left(\frac{jy,K/F}{\nu}\right)
 \prod_{\nu\mbox{ unramified}}\left(\frac{jy,K/F}{\nu}\right)=1.
\label{eq:prod_formula}\end{equation}
The product on the ramified primes yields
$\left(\frac{jy,K/F}{\mathfrak{p}_{7}}\right)
 \left(\frac{jy,K/F}{\mathfrak{q}_{7}}\right)$,
since the ramification in $K/\mathbb{Q}(j)$ is in $7$ only. Note
that
$\left(\frac{xy,K/F}{\nu}\right)=
\left(\frac{x,K/F}{\nu}\right)\left(\frac{y,K/F}{\nu}\right)$
by linearity. We now look at the product on the unramified primes.
Since $y\in\mathfrak{p}_{79}$, its valuation is zero for
$\nu\neq\mathfrak{p}_{79}$.
The valuation of a unit is zero for all places, so that we get
\[
\prod_{\nu\mbox{ unramified}}\left(\frac{jy,K/F}{\nu}\right)
=\prod_{\nu\mbox{ unramified}}\left(\frac{j,K/F}{\nu}\right)
                              \left(\frac{y,K/F}{\nu}\right)
=\left(\frac{y,K/F}{\mathfrak{p}_{79}}\right).
\]
Thus equation (\ref{eq:prod_formula}) simplifies to
\[
\left(\frac{j,K/F}{\mathfrak{p}_{7}}\right)
\left(\frac{y,K/F}{\mathfrak{p}_{7}}\right)
\left(\frac{jy,K/F}{\mathfrak{q}_{7}}\right)
\left(\frac{y,K/F}{\mathfrak{p}_{79}}\right)=1.
\]
The second and third terms are 1 by choice of $y$ (see equations
(\ref{y_cond1}) and (\ref{y_cond2})), so that finally we have
\[
\left(\frac{j,K/F}{\mathfrak{p}_{7}}\right)
\left(\frac{y,K/F}{\mathfrak{p}_{79}}\right)=1.
\]
Since $\mathfrak{p}_{79}$ is inert,
the second term is different
from 1, so that $\left(\frac{j,K/F}{\mathfrak{p}_{7}}\right)\neq1$.
In words, $j$ is not a norm in $\mathfrak{p}_{7}$ which
concludes the proof.
\end{proof}

\begin{prop}\label{prop:j2-not-norm}
The unit $j^2$ is not a norm in
$K/F=\QQ(j,2\cos\left(\frac{2\pi}{7}\right))/\QQ(j)$.
\end{prop}
\begin{proof}
The proof that $j^2$ is not a norm is similar to the above one.  We
keep the notation of the above proof. We show that $j^2$ is not a
norm locally in $\mathfrak{p}_{7}$,
thus $j^2$ is not a norm in $K$.\\
Let $y=5j-9$. We have that
\begin{eqnarray}
y    & \equiv & 1~(\mbox{mod }j-2)\label{y2_cond1}\\
j^2y & \equiv & 1~(\mbox{mod }3+j)\label{y2_cond2}
\end{eqnarray}
and $(y)\ZZ[j]=\mathfrak{p}_{151}$.
Repeating the same computations as in the above proof, we get
\[
\left(\frac{j^2,K/F}{\mathfrak{p}_{7}}\right)
\left(\frac{y,K/F}{\mathfrak{p}_{151}}\right)=1,
\]
where $\mathfrak{p}_{151}$ is inert.
This implies that $j^2$ is not a norm.
\end{proof}

%
%

\section{$i^2=-1$ is not a norm in
$\QQ(i,2\cos(\frac{2\pi}{15}))/\QQ(i)$.}\label{sec:-1-not-norm-4}

We prove here that $i^2=-1$ is not a norm in
$\QQ(i,2\cos(\frac{2\pi}{15}))/\QQ(i)$. The general scheme of the proof is the
same as in Appendix \ref{sec:j-not-norm-3},
though we have to be a bit more careful here,
since the ramification in $\QQ(i,2\cos(\frac{2\pi}{15}))/\QQ(i)$
appears in two primes, unlike in $\QQ(j,2\cos\left(\frac{2\pi}{7}\right))/\QQ(j)$.

\begin{prop}
The unit -1 is not a norm in
$K/F=\QQ(i,2\cos(\frac{2\pi}{15}))/\QQ(i)$.
\end{prop}
\begin{proof}
We consider the field extension $K/F$. We have
\[
5\ZZ[i]=(i+2)(i-2)=\mathfrak{p}_5\mathfrak{q}_5 \mbox{ and }
3\ZZ[i]=3= \mathfrak{p}_3.
\]
We show that $i$ is not a norm locally in $\mathfrak{p}_5$,
thus $i$ is not a norm in $K$.
We look for a number $y$ in $\ZZ[i]$ satisfying
\begin{eqnarray}
y  & \equiv & 1~(\mbox{mod }i+2)\label{y3_cond1}\\
-y & \equiv & 1~(\mbox{mod }i-2)\label{y3_cond2} \\
-y & \equiv & 1~(\mbox{mod }3)\label{y3_cond3}
\end{eqnarray}
By applying the Chinese Remainder Theorem over $\ZZ[i]$, we find
$y=12i-25$ with  $(y)\ZZ[j]=\mathfrak{p}_{769}$. Let $\left(\frac{x,K/F}{\nu}\right)$
denote the Hasse norm symbol. By the product formula
\begin{equation}
\prod_{\nu}\left(\frac{-y,K/F}{\nu}\right)=1.
\label{eq:prod_formula2}
\end{equation}
The product on the ramified primes yields
$\left(\frac{-y,K/F}{\mathfrak{p}_5}\right)
 \left(\frac{-y,K/F}{\mathfrak{q}_5}\right)
 \left(\frac{-y,K/F}{\mathfrak{p}_3}\right)$,
since the ramification in $K/F$ is only in $5$ and $3$. Since
$y\in\mathfrak{p}_{769}$, its valuation is zero for
$\nu\neq\mathfrak{p}_{769}$. The valuation of a unit is zero for all
places, so that we get for the product on the unramified primes
\[
\prod_{\nu\mbox{ unramified}}\left(\frac{-y,K/F}{\nu}\right)
=\prod_{\nu\mbox{ unramified}}\left(\frac{-1,K/F}{\nu}\right)
                              \left(\frac{y,K/F}{\nu}\right)
=\left(\frac{y,K/F}{\mathfrak{p}_{769}}\right).
\]
Thus equation (\ref{eq:prod_formula2}) simplifies to
\[
\left(\frac{-y,K/F}{\mathfrak{p}_3}\right)
\left(\frac{y,K/F}{\mathfrak{p}_{5}}\right)
\left(\frac{-1,K/F}{\mathfrak{p}_{5}}\right)
\left(\frac{-y,K/F}{\mathfrak{q}_{5}}\right)
\left(\frac{y,K/F}{\mathfrak{p}_{769}}\right)=1.
\]
The first, second and fourth terms are 1 by choice of $y$ (see equations
(\ref{y3_cond1}), (\ref{y3_cond2}) and (\ref{y3_cond3})), so that finally
we have
\[
\left(\frac{-1,K/F}{\mathfrak{p}_5}\right)
\left(\frac{y,K/F}{\mathfrak{p}_{769}}\right)=1.
\]
Since $\mathfrak{p}_{769}$ does not split completely,
the second term is different
from 1, so that $\left(\frac{-1,K/F}{\mathfrak{p}_5}\right)\neq1$, which
concludes the proof.

\end{proof}

%
%

\section{$(-j)^3=-1$ is not a norm~ in
$\QQ(j,\zeta_{28}+\zeta_{28}^{-1})/\QQ(j)$.}\label{sec:-1-not-norm-6}

We prove here that $(-j)^3=-1$ is not a norm in
$\QQ(j,\zeta_{28}+\zeta_{28}^{-1})/\QQ(j)$. The proof is similar to
that of Appendix \ref{sec:-1-not-norm-4}.

\begin{prop}
The unit -1 is not a norm in
$K/F=\QQ(\zeta_{28}+\zeta_{28}^{-1},j)/\QQ(j)$.
\end{prop}
\begin{proof}
We consider the field extension $K/F$. We have
\[
7\ZZ[j]=(j-2)(j+3)=\mathfrak{p}_7\mathfrak{q}_7 \mbox{ and }
2\ZZ[j]=2= \mathfrak{p}_2.
\]
We show that $-1$ is not a norm locally in $\mathfrak{p}_7$,
thus $-1$ is not a norm in $K$.

We look for a number $y$ in $\ZZ[j]$ satisfying
\begin{eqnarray}
y  & \equiv & 1~(\mbox{mod }j-2)\label{y4_cond1}\\
-y & \equiv & 1~(\mbox{mod }3+j)\label{y4_cond2} \\
-y & \equiv & 1~(\mbox{mod }2)\label{y4_cond3}
\end{eqnarray}
By applying the Chinese Remainder Theorem over $\ZZ[j]$, we find $y=3-8j$ with
$(y)\ZZ[j]=\mathfrak{p}_{97}$. Let $\left(\frac{x,K/F}{\nu}\right)$
denote the Hasse norm symbol. By the product formula
\begin{equation}
\prod_{\nu}\left(\frac{-y,K/F}{\nu}\right)=1.
\label{eq:prod_formula3}
\end{equation}
The product on the ramified primes yields
$\left(\frac{-y,K/F}{\mathfrak{p}_7}\right)
 \left(\frac{-y,K/F}{\mathfrak{q}_7}\right)
 \left(\frac{-y,K/F}{\mathfrak{p}_2}\right)$,
since the ramification in $K/F$ is in $7$ and $2$ only. Since
$y\in\mathfrak{p}_{97}$, its valuation is zero for
$\nu\neq\mathfrak{p}_{97}$. The valuation of a unit is zero for all
places, so that we get for the product on the unramified primes
\[
\prod_{\nu\mbox{ unramified}}\left(\frac{-y,K/F}{\nu}\right)
=\prod_{\nu\mbox{ unramified}}\left(\frac{-1,K/F}{\nu}\right)
                              \left(\frac{y,K/F}{\nu}\right)
=\left(\frac{y,K/F}{\mathfrak{p}_{97}}\right).
\]
Thus equation (\ref{eq:prod_formula3}) simplifies to
\[
\left(\frac{-y,K/F}{\mathfrak{p}_2}\right)
\left(\frac{y,K/F}{\mathfrak{p}_7}\right)
\left(\frac{-1,K/F}{\mathfrak{p}_7}\right)
\left(\frac{-y,K/F}{\mathfrak{q}_7}\right)
\left(\frac{y,K/F}{\mathfrak{p}_{97}}\right)=1.
\]
The first, second and fourth terms are 1 by choice of $y$ (see equations
(\ref{y4_cond1}), (\ref{y4_cond2}) and (\ref{y4_cond3})), so that finally
we have
\[
\left(\frac{-1,K/F}{\mathfrak{p}_7}\right)
\left(\frac{y,K/F}{\mathfrak{p}_{97}}\right)=1.
\]
Since $\mathfrak{p}_{97}$ does not split completely,
the second term is different
from 1, so that $\left(\frac{-1,K/F}{\mathfrak{p}_7}\right)\neq1$, which
concludes the proof.

\end{proof}

%
%

\section{The LLL reduction algorithm over $\ZZ[j]$}\label{sec:LLL}

The standard LLL reduction algorithm \cite{Grotschel}
over $\ZZ$ can be easily modified to work over $\ZZ[j]$ \cite{Huguette}.
The two main points to be careful about are
\begin{itemize}
\item
the Euclidean division: the quotient of the Euclidean division
over $\ZZ[j]$ is defined as follows: let $x=x_1+ j x_2$ and
$y=y_1 + j y_2$, $x_1,x_2,y_1,y_2 \in \ZZ$. The division of
$x$ by $y$ yields $\frac{x}{y}= z_1 + j z_2$, with $z_1,z_2 \in \QQ$.
Then we have that $x = yq+r$, where $q=[z_1]+j[z_2]$.
\item the conjugation: the usual complex conjugation is replaced by
the $\tau$-conjugation, that sends $j$ onto $j^2$.
\end{itemize}

%
%

%
%
\newpage
~\\

\listoffigures

%
%

\begin{figure}[p]
\[
\begin{diagram}
\node[2]{K} \arrow{sw,t,-}{n} \arrow[2]{s,b,-}{2n} \arrow{se,t,-}{2}\\
\node{F} \arrow{se,b,-}{2} \node[2]{\QQ(\theta)} \arrow{sw,b,-}{n} \\
\node[2]{\QQ}
\end{diagram}
\]
\caption{\label{fig:comp}
The compositum of a totally real field $\QQ(\theta)$ and
$F=\QQ(i)$ or $\QQ(j)$ with coprime discriminants:
relative degrees are shown on the branches.}
\end{figure}

\begin{figure}[p]
\begin{center}
\includegraphics[width=10cm,keepaspectratio]{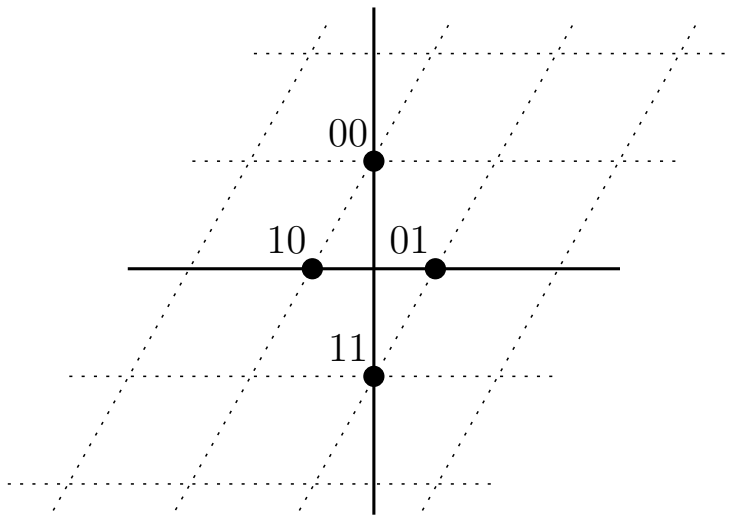}
\end{center}
\caption{\label{cap:Hex-const-4} 4-HEX Constellation.}
\end{figure}

\begin{figure}[p]
\begin{center}
\includegraphics[width=13cm,keepaspectratio]{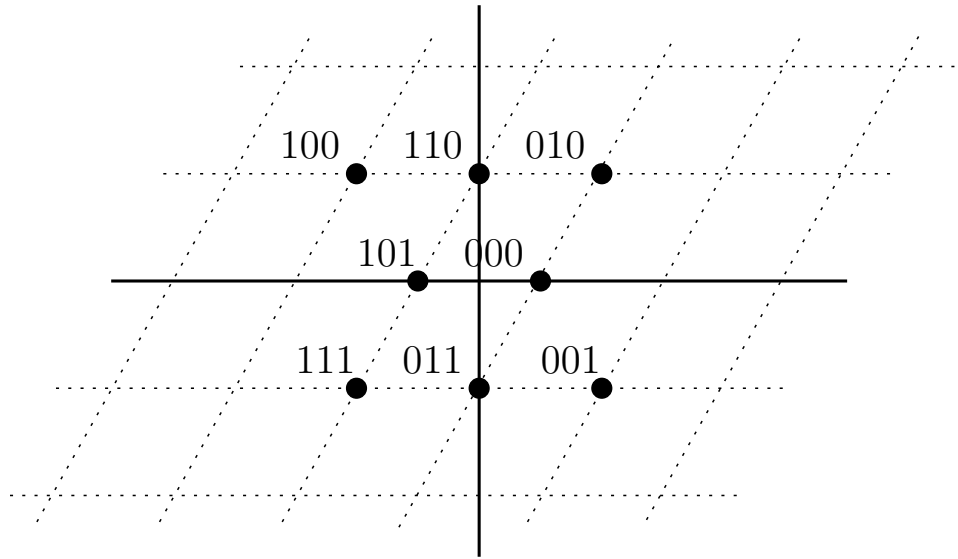}
\end{center}
\caption{\label{cap:Hex-const-8} 8-HEX Constellation.}
\end{figure}

\begin{figure}[p]
\begin{center}
\includegraphics[width=14cm,keepaspectratio]{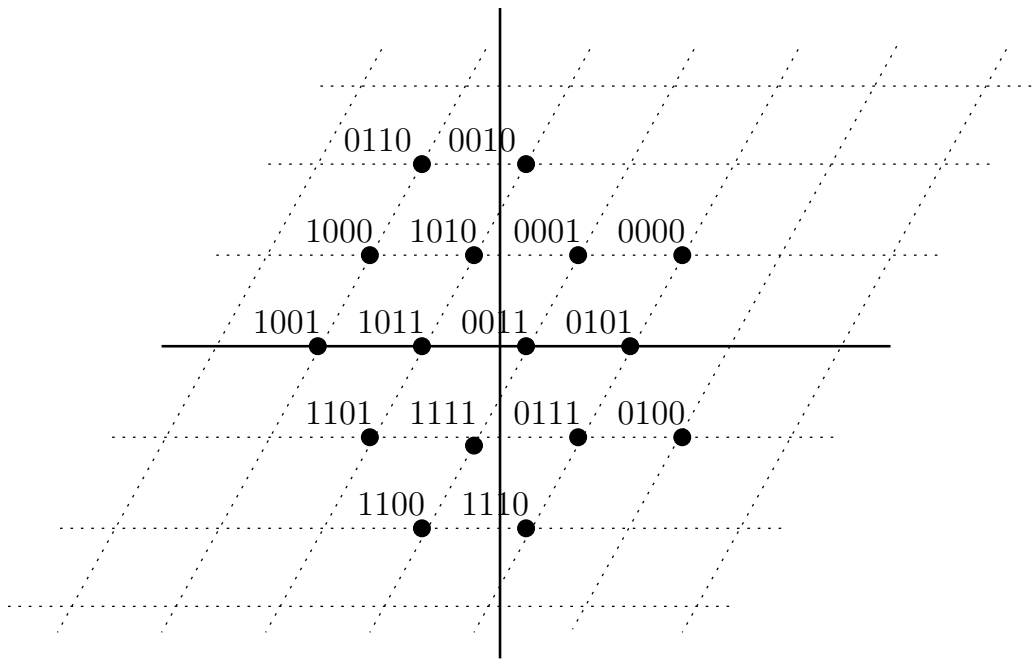}
\end{center}
\caption{\label{cap:Hex-const-16} 16-HEX Constellation.}
\end{figure}

\begin{figure}[p]
\begin{center}
\includegraphics[width=16cm,keepaspectratio]{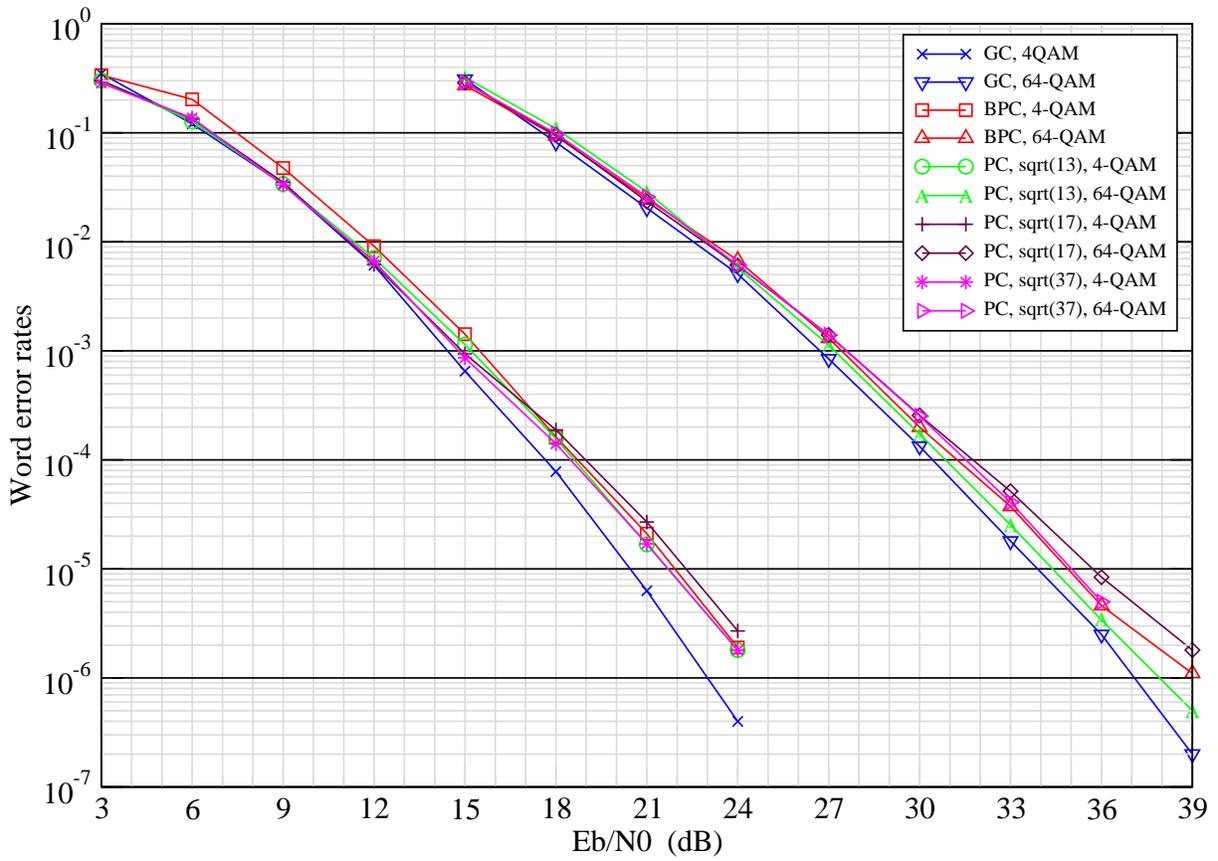}
\end{center}
\caption{\label{cap:fig22}
Golden code (GC) (perfect code with $p=5$) and other perfect
codes (PC) compared to the best peviously known $2 \times 2$ codes (BPC).}
\end{figure}

\begin{figure}[p]
\begin{center}
\includegraphics[width=16cm,keepaspectratio]{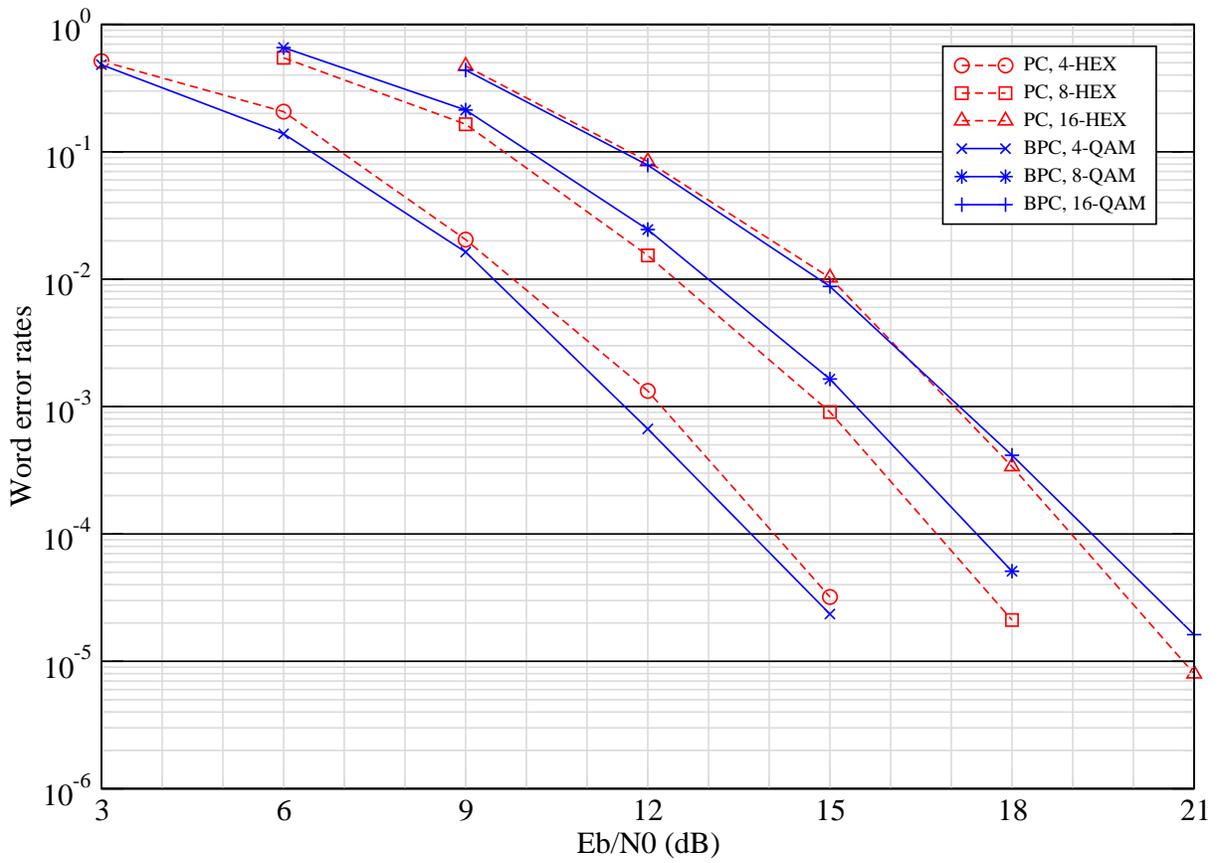}
\end{center}
\caption{\label{cap:fig33}
Perfect $3 \times 3$ codes with HEX symbols (PC) compared
to the best peviously known codes (BPC).}
\end{figure}

\begin{figure}[p]
\begin{center}
\includegraphics[width=16cm,keepaspectratio]{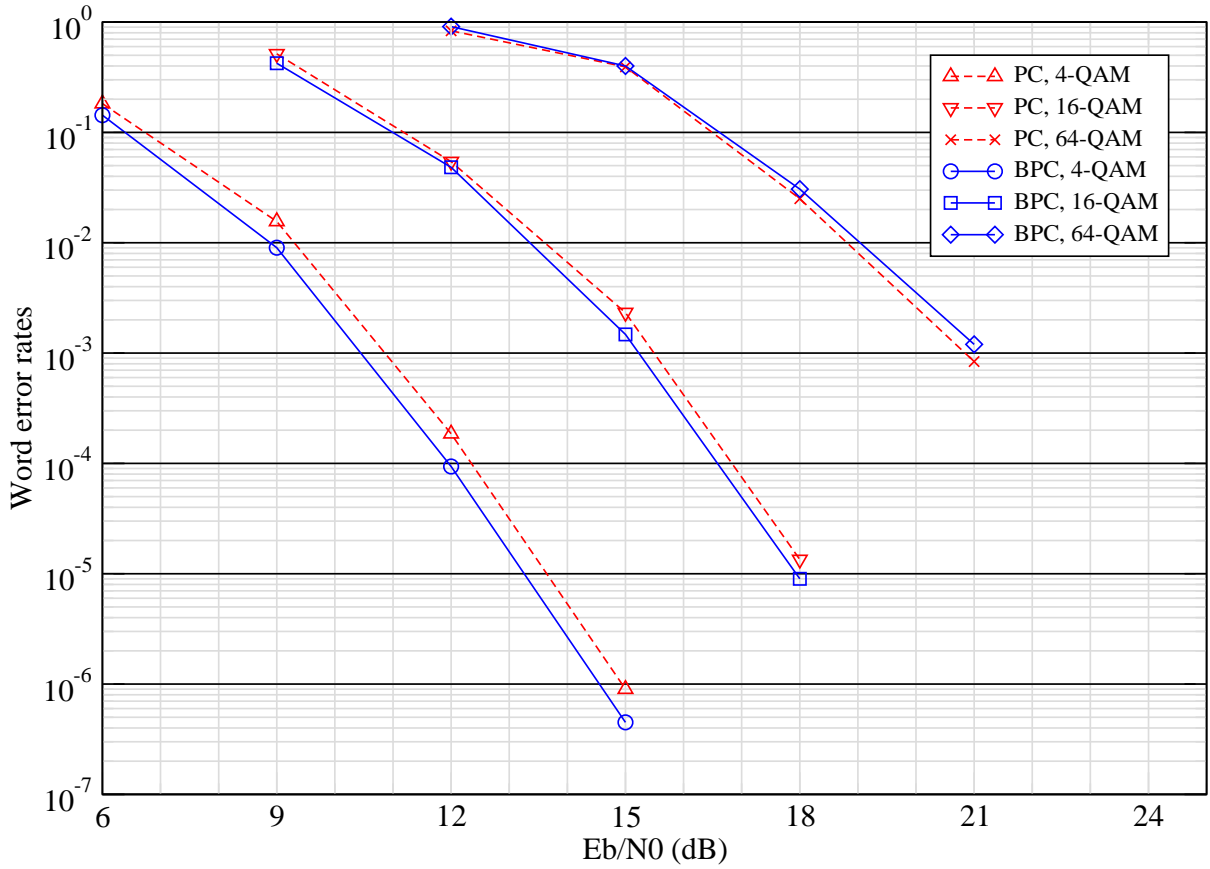}
\end{center}
\caption{\label{cap:fig44}
Perfect $4 \times 4$ codes with QAM symbols (PC) compared to
the best peviously known codes (BPC).}
\end{figure}

\end{document}